\documentclass[10pt]{article}
\usepackage{amssymb,amsmath}
\usepackage{amsthm}
\usepackage{mathrsfs}
\usepackage{dcolumn}
\usepackage{bm}
\usepackage{fullpage}
\usepackage{color}
\usepackage[all]{xy}
\usepackage[dvipdfmx]{graphicx,hyperref}
\usepackage{ulem}
\usepackage{ascmac}
\usepackage{caption}
\usepackage{here}
\usepackage{comment}

\begin{document}
\theoremstyle{definition}
\newtheorem{Definition}{Definition}[section]
\newtheorem{Theorem}[Definition]{Theorem}
\newtheorem*{theorem}{Theorem}
\newtheorem{Proposition}[Definition]{Proposition}
\newtheorem{Question}[Definition]{Question}
\newtheorem{Lemma}[Definition]{Lemma}
\newtheorem*{Proof}{Proof}
\newtheorem{Example}[Definition]{Example}
\newtheorem{Postulate}[Definition]{Postulate}
\newtheorem{Corollary}[Definition]{Corollary}
\newtheorem{Remark}[Definition]{Remark}
\newtheorem{Claim}[Definition]{Claim}
\newtheorem{Assumption}[Definition]{Assumption}

\theoremstyle{remark}
\newcommand{\beq}{\begin{equation}}
\newcommand{\beqa}{\begin{eqnarray}}
\newcommand{\eeq}{\end{equation}}
\newcommand{\eeqa}{\end{eqnarray}}
\newcommand{\non}{\nonumber}
\newcommand{\lb}{\label}
\newcommand{\fr}[1]{(\ref{#1})}
\newcommand{\bb}{\mbox{\boldmath {$b$}}}
\newcommand{\bbe}{\mbox{\boldmath {$e$}}}
\newcommand{\bt}{\mbox{\boldmath {$t$}}}
\newcommand{\bn}{\mbox{\boldmath {$n$}}}
\newcommand{\br}{\mbox{\boldmath {$r$}}}
\newcommand{\bC}{\mbox{\boldmath {$C$}}}
\newcommand{\bp}{\mbox{\boldmath {$p$}}}
\newcommand{\bx}{\mbox{\boldmath {$x$}}}
\newcommand{\bF}{\mbox{\boldmath {$F$}}}
\newcommand{\bT}{\mbox{\boldmath {$T$}}}
\newcommand{\bQ}{\mbox{\boldmath {$Q$}}}
\newcommand{\bS}{\mbox{\boldmath {$S$}}}
\newcommand{\balpha}{\mbox{\boldmath {$\alpha$}}}
\newcommand{\bomega}{\mbox{\boldmath {$\omega$}}}
\newcommand{\ve}{{\varepsilon}}
\newcommand{\e}{\mathrm{e}}
\newcommand{\f}{\mathrm{f}}
\newcommand{\s}{\mathrm{s}}
\newcommand{\B}{\mathrm{B}}
\newcommand{\E}{\mathrm{E}}
\newcommand{\G}{\mathrm{G}}
\newcommand{\R}{\mathrm{R}}
\newcommand{\Z}{\mathrm{Z}}
\newcommand{\HH}{\mathrm{H}}
\newcommand{\I}{\mathrm{I}}
\newcommand{\II}{\mathrm{II}}
\newcommand{\Id}{\mathrm{Id}}
\newcommand{\ddiv}{\mathrm{div}}
\newcommand{\Ising}{\footnotesize\mathrm{Ising}}
\newcommand{\std}{\mathrm{std}}
\newcommand{\hF}{\widehat F}
\newcommand{\hL}{\widehat L}
\newcommand{\tA}{\widetilde A}
\newcommand{\tB}{\widetilde B}
\newcommand{\tC}{\widetilde C}
\newcommand{\tL}{\widetilde L}
\newcommand{\tK}{\widetilde K}
\newcommand{\tX}{\widetilde X}
\newcommand{\tY}{\widetilde Y}
\newcommand{\tU}{\widetilde U}
\newcommand{\tZ}{\widetilde Z}
\newcommand{\talpha}{\widetilde \alpha}
\newcommand{\te}{\widetilde e}
\newcommand{\tv}{\widetilde v}
\newcommand{\ts}{\widetilde s}
\newcommand{\tx}{\widetilde x}
\newcommand{\ty}{\widetilde y}
\newcommand{\ud}{\underline{\delta}}
\newcommand{\uD}{\underline{\Delta}}
\newcommand{\chN}{\check{N}}
\newcommand{\cA}{{\cal A}}
\newcommand{\cB}{{\cal B}}
\newcommand{\cC}{{\cal C}}
\newcommand{\cD}{{\cal D}}
\newcommand{\cE}{{\cal E}}
\newcommand{\cF}{{\cal F}}
\newcommand{\cG}{{\cal G}}
\newcommand{\cH}{{\cal H}}
\newcommand{\cI}{{\cal I}}
\newcommand{\cJ}{{\cal J}}
\newcommand{\cK}{{\cal K}}
\newcommand{\cL}{{\cal L}}
\newcommand{\cM}{{\cal M}}
\newcommand{\cN}{{\cal N}}
\newcommand{\cO}{{\cal O}}
\newcommand{\cP}{{\cal P}}
\newcommand{\cQ}{{\cal Q}}
\newcommand{\cS}{{\cal S}}
\newcommand{\cT}{{\cal T}}
\newcommand{\cY}{{\cal Y}}
\newcommand{\cU}{{\cal U}}
\newcommand{\cV}{{\cal V}}
\newcommand{\cW}{{\cal W}}
\newcommand{\vecX}{\mathfrak{X}}
\newcommand{\tcA}{\widetilde{\cal A}}
\newcommand{\DD}{{\cal D}}
\newcommand\TYPE[3]{ \underset {(#1)}{\overset{{#3}}{#2}}  }
\newcommand{\Qc}{\overset{\footnotesize\circ}{Q}}
\newcommand{\bfe}{\boldsymbol e} 
\newcommand{\bfb}{{\boldsymbol b}}
\newcommand{\bfd}{{\boldsymbol d}}
\newcommand{\bfh}{{\boldsymbol h}}
\newcommand{\bfj}{{\boldsymbol j}}
\newcommand{\bfn}{{\boldsymbol n}}
\newcommand{\bfA}{{\boldsymbol A}}
\newcommand{\bfB}{{\boldsymbol B}}
\newcommand{\bfJ}{{\boldsymbol J}}
\newcommand{\bfS}{{\boldsymbol S}}
\newcommand{\dr}{\,\mathrm{d}}
\newcommand{\Dr}{\mathrm{D}}
\newcommand{\saddle}{\mathrm{saddle}}
\newcommand{\can}{\mathrm{can}}
\newcommand{\const}{\mathrm{const.}}
\newcommand{\eq}{\,\mathrm{eq}}
\newcommand{\wt}[1]{\widetilde{#1}}
\newcommand{\wh}[1]{\widehat{#1}}
\newcommand{\ch}[1]{\check{#1}}
\newcommand{\ol}[1]{\overline{#1}}
\newcommand{\ii}{\imath}
\newcommand{\ic}{\iota}
\newcommand{\mbbP}{\mathbb{P}}
\newcommand{\mbbR}{\mathbb{R}}
\newcommand{\mbbN}{\mathbb{N}}
\newcommand{\mbbZ}{\mathbb{Z}}
\newcommand{\Leftrightup}[1]{\overset{\mathrm{#1}}{\Longleftrightarrow}}
\newcommand{\avg}[1]{\left\langle\,{#1}\, \right\rangle}
\newcommand{\step}{\lrcorner\hspace*{-0.55mm}\ulcorner}
\newcommand{\equp}[1]{\overset{\mathrm{#1}}{=}}
\newcommand{\nequp}[1]{\overset{\mathrm{#1}}{\tiny\neq}}
\newcommand{\equpg}[2]{\overset{\mathrm{#2}}{#1}}
\newcommand{\nin}{\in\hspace{-.78em}\setminus}
\newcommand{\fraX}{\mathfrak{X}}
\newcommand{\Gam}[1]{\Gamma{#1}}
\newcommand{\GamLamM}[1]{{\Gamma\Lambda^{{#1}}\cal{M}}}
\newcommand{\GamLam}[2]{{\Gamma\Lambda^{{#2}}{#1}}}
\newcommand{\GTM}{\Gamma T{\cal M}}
\newcommand{\GT}[1]{{\Gamma T{#1}}}
\newcommand{\inp}[2]{\left\langle\,  #1\, , \, #2\, \right\rangle}
\newcommand{\inpr}[2]{\left(\,  #1\, , \, #2\, \right)}
\newcommand{\rmt}[1]{{\mathrm{t}}({#1})}
\newcommand{\rmo}[1]{{\mathrm{o}}({#1})}
\newcommand{\paral}{/\hspace*{-2pt}/}

\title{Contact topology and electromagnetism:\\
The Weinstein conjecture and  Beltrami-Maxwell fields}
\author{\large Shin-itiro GOTO 
\\
Center for  Mathematical Science and Artificial Intelligence,\\
Chubu University,\quad 
1200 Matsumoto-cho, Kasugai, Aichi 487-8501, Japan
}
\date{}
\maketitle
\begin{abstract}%
  We draw connections between contact topology 
  and Maxwell fields  
   in vacuo on $3$-dimensional closed Riemannian submanifolds 
  in $4$-dimensional Lorentzian manifolds. 
  This is accomplished by showing that
  contact topological methods can be applied to reveal 
  topological features of a class of solutions to Maxwell's equations. 
  This class of Maxwell fields is 
  such that electric fields are parallel to
  magnetic fields. In addition these electromagnetic fields
  are composed of the so-called Beltrami fields.   
  We employ several theorems resolving   
  the Weinstein conjecture on closed
  manifolds with contact structures and stable Hamiltonian structures,
  where this conjecture refers to the 
  existence of periodic orbits of the Reeb vector fields.  
  Here a contact form is a special case of a stable Hamiltonian structure.
  After showing how to relate Reeb vector fields with electromagnetic
  $1$-forms,  we apply a theorem regarding 
  contact manifolds
  and an improved  theorem regarding stable Hamiltonian structures.   
  Then a closed field line is shown to exist, where field lines are generated by
  Maxwell fields. 
  In addition, 
  electromagnetic energies are shown to be conserved along the
  Reeb vector fields. 
  
\end{abstract}%


\section{Introduction}
\label{section-introduction}
Maxwell's equations are the most fundamental equations in electromagnetism, 
and they describe electromagnetic phenomena in the form of field equations.  
These equations have been studied in  mathematical sciences, and have been
applied to engineering problems\,\cite{Jackson1998}. Despite its long history, 
various problems associated with 
Maxwell's equations are still actively studied, 
and its theoretical progress is expected to promote the development of
mathematical theories and engineering applications further\,\cite{Heal2003}.
One important class of the Maxwell fields is  
such that electric fields are parallel to magnetic fields,
and this class of fields 
is expected to apply to plasma heating and
acceleration. See \cite{Lakhtakia1994,Mochizuki2022} and references therein
for its brief history of the applications of these 
electromagnetic fields.    
From the mathematical side, various elegant formulations of Maxwell's equations 
have been proposed.
One of such is based on 
pseudo Riemannian geometry, and this is a standard description
in theoretical physics\,\cite{Frenkel,Benn1987,Abraham1998,Burton2024}. 
In this description $2$-forms on $4$-dimensional manifolds play 
pivotal roles, where manifolds are trivial bundles of the form 
$\mbbR\times\cN$. 
This $\mbbR$ expresses time physically,  
and $\cN$ is a $3$-dimensional manifold expressing physical space. 
Several ideas and notions known in pseudo Riemannian geometry 
have been applied to various problems in solving
Maxwell's equations\,\cite{Tucker2007,Tucker2022,Goto2009}. 
Aside from considerable 
progress in understanding mathematics regarding Maxwell's equations  
in terms of pseudo Riemannian geometry, 
there is a possibility to develop new theories in terms of other geometries 
to understand Maxwell's equations on manifolds, and such developments  
are expected to shed light on other aspects of Maxwell's equations.

Recall that geometry for $4,6,\ldots$-dimensional manifolds is often
associated with symplectic geometry, 
and that geometry for $3,5,\ldots$-dimensional manifolds is 
often associated with contact geometry\,\cite{McDuff,Silva2008}.
These even and odd dimensional geometries are linked. To be precise,  
a $(k+1)$-dimensional symplectic manifold is constructed from 
a $k$-dimensional contact manifold by means of the so-called symplectization,  
where $k$ is an odd natural number.  
Since symplectic and contact geometries are well-developed branches
of mathematics,  
one conceives the idea that the study of symplectic and contact geometry can
be applied to the study of Maxwell's equations and related areas.
This is indeed the case, and a variety of papers have shown 
the relevance of applying symplectic and contact
geometries\,\cite{Dahl2004,Dahl2008,Gracia2022}. 
Meanwhile, since the paper by Gromov\,\cite{Gromov1985}, 
topological aspects of symplectic and contact geometries  
have been investigated\,\,\cite{Hofer1994}. 
These research activities have flourished into an area of mathematics,
called  symplectic (and contact) topology, where  
symplectic and contact topology combines ideas
mainly from dynamical systems theory
and topology\,\cite{Hofer1994}.
One of the major advances in contact topology is the resolution of  
the Weinstein conjecture, where this resolution is to 
prove the existence of periodic orbits of the so-called
Reeb vector fields on closed contact
manifolds\,\cite{Taubes2007,Hutchings2010}.   
An application of contact topology was then found in fluid mechanics. 
In particular,  in the study of the ideal fluid in  
$3$-dimensional manifolds, some existence
theorems were established, where the so-called
Beltrami field was focused\,\cite{Etnyre2000}. 
Although there are other physical sciences to which contact topology
has been applied\,\cite{Moreno2022,Entov2023},
an explicit application of this topology to electromagnetism
is not yet explored. 
In other words, contact and symplectic topological aspects 
in electromagnetism have been invisible so far. Note that there are
other topological approaches to electromagnetism in the
literature\,\cite{Nakata2019,Gratus2020}. 
We then feel that well-developed symplectic and contact topology
should be applied to the study of Maxwell's equations. 

We aim at drawing connections between 
contact topology and the study of Maxwell's equations, and 
this contribution is a step toward the establishment of a contact
topological method in electromagnetism. 
To avoid various complications, 
in this paper,
a special class of Maxwell's equations is focused. 
This class of Maxwell fields is such that 
electromagnetic vector fields are parallel to 
magnetic fields,
where fields are composed of the so-called Beltrami fields.
As mentioned above, this class of Maxwell's equations is important in
physical applications. In addition, this class is closely related to 
the ideal fluid in the sense that both of velocity fields for ideal fluid and 
electromagnetic fields are constructed from 
the Beltrami field. Because the ideal fluid described by the Beltrami field
has been well-studied mathematically, some of existing statements
are expected to be employed for the study of this class of Maxwell's equations.
Advanced methods that we employ in contact topology are 
several theorems resolving the Weinstein conjecture. 
In particular, we employ Taubes's proof for the Weinstein conjecture in
$3$-dimensional contact manifolds and the Hutchings and Taubes proof
for stable Hamiltonian structures in $3$-dimensional 
manifolds\,\cite{Taubes2007,Hutchings2009}. 

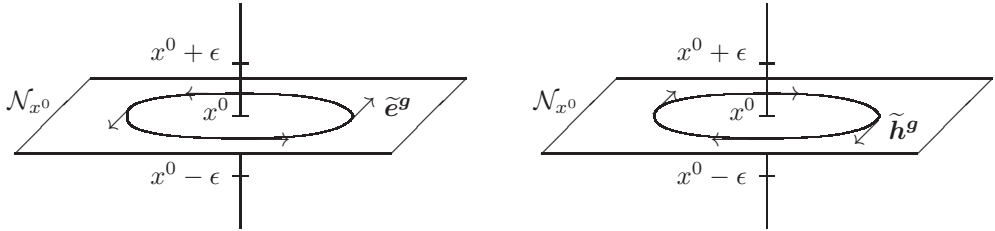
\begin{figure}[htb]
\begin{picture}(120,80)
\unitlength 1mm
\put(40,14.5){$x^{0}$}
\put(14,16){$\cN_{x^{0}}$}
\put(45,15){\line(0,1){15}}
\put(44,22){\line(1,0){2}}
\put(44,15){\line(1,0){2}}
\put(44,7){\line(1,0){2}}
\put(33,22.5){$x^{0}+\epsilon$}
\put(33,6){$x^{0}-\epsilon$}
\put(45,0){\line(0,1){10}}
\put(15,10){\line(1,0){50}}
\put(25,20){\line(1,0){50}}
\put(15,10){\line(1,1){10}}
\put(65,10){\line(1,1){10}}
\qbezier(45,18)(58.5,18)(60,15)
\qbezier(45,12)(58.5,12)(60,15)
\qbezier(45,18)(29,18)(30,15)
\qbezier(45,12)(29,12)(30,15)
\put(59.4,15){$\nearrow$\ $\wt{\bm{e}}^{\bm{g}}$}
\put(27.3,14){$\swarrow$} 
\put(46,11.1){$\longrightarrow$}
\put(37.5,17.1){$\leftarrow$}
\put(110,14.5){$x^{0}$}
\put(84,16){$\cN_{x^{0}}$}
\put(115,15){\line(0,1){15}}
\put(114,22){\line(1,0){2}}
\put(114,15){\line(1,0){2}}
\put(114,7){\line(1,0){2}}
\put(103,22.5){$x^{0}+\epsilon$}
\put(103,6){$x^{0}-\epsilon$}
\put(115,0){\line(0,1){10}}
\put(85,10){\line(1,0){50}}
\put(95,20){\line(1,0){50}}
\put(85,10){\line(1,1){10}}
\put(135,10){\line(1,1){10}}
\qbezier(115,18)(128.5,18)(130,15)
\qbezier(115,12)(128.5,12)(130,15)
\qbezier(115,18)(100,18)(100,15)
\qbezier(115,12)(100,12)(100,15)
\put(126.5,12){$\swarrow$\ $\wt{\bm{h}}^{\bm{g}}$}
\put(99.6,15.9){$\nearrow$}
\put(116,17.1){$\rightarrow$}
\put(107.5,11.1){$\longleftarrow$}
\linethickness{0.3mm}
\end{picture}
\caption{Closed field lines of $\bm{e}$ and $\bm{h}$ at $x^{0}$,
  where $\epsilon>0$ is a small number. 
  These field lines are integral curves of
  the vector fields $\wt{\bm{e}}^{\bm{g}}$ and $\wt{\bm{h}}^{\bm{g}}$.
  Note that in the main theorems,
  $\cN_{x^{0}}$ must be closed, where a closed manifold
  is a manifold without boundary that is compact. }
\label{picture-field-lines-e-h}
\end{figure}

This article concentrates attention on Maxwell's equations in vacuo, where 
these equations are described in the form language on manifolds,
and these solutions 
are called Maxwell fields. They are explained as follows. 
Let $(\mbbR\times\cN,g)$ be a Lorentzian manifold, where 
$\cN$ is a $3$-dimensional manifold and $g=-\dr x^{0}\otimes\dr x^{0}+\bm{g}$ 
with $x^{0}$ being the coordinate of $\mbbR$.
This $\bm{g}$ is a Riemannian metric on $\cN$, 
and does not depend on $x^{0}$ in the sense of 
$\cL_{\partial/\partial x^{0}}\bm{g}=0$, where 
$\cL_{X}$ denotes the Lie derivative along a vector field $X$.  
At a fixed $x^{0}$, $\cN_{x^{0}}$ denotes the hypersurface in
$\mbbR\times \cN$, and $(\cN_{x^{0}},\bm{g})$ becomes a Riemannian manifold.
This $\cN_{x^{0}}$ is abbreviated by simply $\cN$ if there is no risk of
confusion. 
Let $\bm{e}$ and $\bm{h}$ be $1$-forms and put 
$\bm{B}=\mu_{0}\star_{3}\bm{h},\bm{D}=\varepsilon_{0}\star_{3}\bm{e}$,
where $\varepsilon_{0},\mu_{0}$ are constant and
$\star_{3}$ is the Hodge map associated with $\bm{g}$.    
In addition, let these forms, $\bm{e},\bm{h},\bm{B},\bm{D}$,  
solve Maxwell's equations 
in vacuo,
$\bm{\dr}\bm{e}=-\cL_{\partial/\partial t}\bm{B}$, $\bm{\dr}\bm{B}=0$,
$\bm{\dr}\bm{D}=0$, $\bm{\dr}\bm{h}=\cL_{\partial/\partial t}\bm{D}$, where
$t=x^{0}/c_{0}$ with $c_{0}$ denoting the speed of light in vacuo. 
In this article attention is restricted to the field lines of
electromagnetic fields at a fixed $x^{0}=c_{0}t$,  
where field lines are integral curves of the vector fields
$\wt{\bm{e}}^{\bm{g}}:=\bm{g}^{-1}(\bm{e},-)$ and
$\wt{\bm{h}}^{\bm{g}}:=\bm{g}^{-1}(\bm{h},-)$ on
$\cN_{x^{0}}$    
(see Figure\,\ref{picture-field-lines-e-h}).

In the physics literature, 
these field lines have been considered in numerous systems.  
To be precise, electromagnetic field lines at a fixed $x^{0}$ are often 
depicted to show how fields are spatially distributed. 
For instance, place a point charge whose electric charge is $q>0$ at 
the origin $(0,0,0)$  
in the Riemannian manifold $(\mbbR^{3},\bm{g})$ with  
$\bm{g}=\dr x^{1}\otimes \dr x^{1}+\dr x^{2}\otimes \dr x^{2}+\dr x^{3}\otimes \dr x^{3}$. Then the electric field generated by this charge
is written in the form language as 
$\bm{e}=q/(4\pi\varepsilon_{0}r^{2})\dr r$ with
$r=\sqrt{x^{2}+y^{2}+z^{2}}$ for
an observer at rest. The corresponding field lines are 
straight lines that  
radiate outward from the origin $(0,0,0)$ in $\mbbR^{3}$. 
Although this example only illustrates simple straight field lines, 
field lines in general form non-trivial structures such as
knots\,\cite{Arrayas2017,Bode2021}. 
In these nontrivial cases closed field lines can play a role of
backbones of field configurations, where this role played in 
electromagnetism is the same as the role of closed orbits 
in dynamical systems theory.
  Because of these, closed field lines in electromagnetism should be
  of interest, and be studied.
  To advance our understanding of closed electromagnetic field lines,
  this paper shows the existence of closed field lines theoretically.  
  Meanwhile an experimental realization of an electromagnetic
  knot has been challenging 
  as mentioned in \cite{Arrayas2017}. Since closed electromagnetic field lines  
  are akin to simple electromagnetic knots, a feasible way to realize
  closed electromagnetic field lines in a laboratory setting
  should be sought in future. 
In the mathematics literature 
the so-called magnetic field lines have been considered\,\cite{Inoguchi2017}.

The following is one of main theorems in this paper, and this
is an existence theorem.
\begin{Theorem}[Theorem\,\ref{fact-Taubes-2007-electromagnetism} of this paper]
Fix $x^{0}$, and let $(\cN_{x^{0}},\bm{g})$ 
be a closed oriented connected 
$3$-dimensional Riemannian manifold.  
Let $\bm{B},\bm{D},\bm{e},\bm{h}$ be time-dependent Maxwell fields on
$\cM_{\epsilon}$  
with 
$\bm{e}\wedge\bm{\dr}\bm{e}\neq 0$ and 
$\bm{h}\wedge\bm{\dr}\bm{h}\neq 0$ on $\cN_{x^{0}}$. 
Then there are closed field lines of $\bm{e}$ and $\bm{h}$
on $\cN_{x^{0}}$.
\end{Theorem}
In the above theorem, the product manifold 
$\cM_{\epsilon}:=(x^{0}-\epsilon,x^{0}+\epsilon)\times\cN_{x_{0}^{0}}$
has been introduced with some $\epsilon>0$. 

To improve the theorem above, one introduces stable Hamiltonian structures 
on $3$-dimensional manifolds, where a contact form is a special case of
a stable Hamiltonian structure
(see Definition\,\ref{definition-stable-Hamiltonian-structure}). 
Let $F_{0}$ be a Maxwell $2$-form,
and $\star^{-1} F_{1}$ be
an excitation $2$-form in vacuo, where $\star$ is a Hodge map
associated with $g$. They are $F_{0}=-c_{0}\bm{B}-\bm{e}\wedge\dr x^{0}$ and
$F_{1}=\bm{D}-c_{0}^{-1}\bm{h}\wedge\dr x^{0}$ defined on
$\cM_{\epsilon}$.  
Then the following is shown: 
\begin{Theorem}[Theorem\,\ref{fact-Hutchings2009-electromagnetism} of this paper]
  Fix $x^{0}$, and
  let $\bm{B},\bm{D},\bm{e},\bm{h}$ be time-dependent Maxwell fields on
  $\cM_{\epsilon}$. 
In addition, let $(\cN_{x^{0}},\bm{g})$
be a closed oriented connected
$3$-dimensional Riemannian manifold with stable Hamiltonian
structures $(\bm{B},\bm{e}/\bm{g}^{-1}(\bm{e},\bm{e}))$ for $F_{0}$
and $(\bm{D},\bm{h}/\bm{g}^{-1}(\bm{h},\bm{h}))$ for $F_{1}$
on $\cN_{x^{0}}$. 
If $\cN_{x^{0}}$ is not a $T^{2}$-bundle over $S^{1}$, then the
field lines of $\bm{e}$ and those of $\bm{h}$ on $\cN_{x^{0}}$  
have closed orbits.
\end{Theorem}

Maxwell fields that can be applied to these main theorems are
generated by Beltrami $1$-forms  
(see Theorem\,\ref{fact-twisted-mode-geometry-summery} and 
Lemma\,\ref{fact-explicit-Maxwell-Beltrami-time-dependent}). 
In the theorem regarding stable Hamiltonian structures, 
the Reeb vector fields are written in terms of
electromagnetic $1$-forms. It will be shown that electromagnetic
energies are conserved along these Reeb vector fields
(see Proposition\,\ref{fact-conserved-energies-along-Reeb-SHS-general} for
a class of Maxwell fields with stable Hamiltonian structures, and
Corollary\,\ref{fact-conserved-energies-along-Reeb-SHS-Beltrami} for
particular Maxwell fields with stable Hamiltonian structures). 

This article is organized as follows.
In Section\,\ref{section-preliminaries}
necessary background is briefly summarized, and   
the notation used in this paper is fixed, where such background is
Maxwell's equations on manifolds, symplectic and contact topology,
and geometry related to Beltrami fields. 
In Section\,\ref{section-electromagnetic-Beltrami}
symplectic, contact, and stable Hamiltonian nature of
Maxwell's equations are discussed.
To this end, it is shown how to relate Maxwell fields with
the Reeb vector fields, and shown that
some electromagnetic energies are conserved along the Reeb vector fields.   
Then for the Maxwell fields
generated by the Beltrami field, it is proved that 
closed electromagnetic field lines exist.  
In addition, physical relevance of obtained theorems is briefly discussed. 
In Section\,\ref{section-dicussion-conclusions} 
our study is summarized, 
and some of future works are discussed. 
Section\,\ref{section-appendix} as an appendix shows
derivations of equations appeared 
in the main text. 

\section{Preliminaries} 
\label{section-preliminaries}
The principal task to theoretically
understand how electromagnetic fields behave  
under given conditions is to solve  
Maxwell's equations\,\cite{Jackson1998}, and geometric formulations often 
facilitate analysis of these equations\,\cite{Tucker2007}. 
One of such formulations of  Maxwell's equations employs
$4$-dimensional Lorentzian manifolds\,\cite{Nakahara,Frenkel}, and another one
employs contact and symplectic manifolds\,\cite{Dahl2008}.
Each of them is briefly summarized below, and 
an amalgamation of these existing formulations 
is employed throughout this paper. 
In addition some preliminary statements about Maxwell's equations are also shown
in this section for later sections. 

In what follows underlining geometric objects are introduced, and notations
are fixed. Let $\cM$ be a $4$-dimensional manifold,
and $X$ a vector field on $\cM$.
The space of vector fields on $\cM$ is denoted by $\GTM$, and  
the space of $k$-forms on $\cM$ is denoted by $\GamLamM{k}$, $(k=0,\ldots,4)$. 
The Lie derivative of a tensor field $T$ along $X$ is denoted by $\cL_{X}T$,
and the Cartan formula for a $p$-form $\alpha$ holds, that is, 
$\cL_{X}\alpha=(\dr \ii_{X}+\ii_{X}\dr)\alpha$,
where $\ii_{X}:\GamLamM{k}\to\GamLamM{k-1}$
and $\dr:\GamLamM{k}\to\GamLamM{k+1}$ denote the
interior product with respect to $X\in\GTM$ and
the exterior derivative, respectively.   
Let $t$ be a coordinate for $\cI\subset\mbbR$ called time, and
$c:\cI\to\cM$, $(t\mapsto p(t))$ be a curve. Then a differential of a tensor
field $T$ on $\cM$ with respect to $t$ can be defined and
is denoted by $\cL_{\partial/\partial t}T$ when $t$ is treated as a coordinate of $\cM$. When $\cL_{\partial/\partial t}T$ appears in the sequel, this treatment is implicitly adopted.   
Let $\cN$ be a $3$-dimensional manifold. Then
$\GamLam{\cN}{k}$ denotes the space of $k$-forms on $\cN$, and so on. 
If a vector field $X$ does not vanish anywhere on some domain
under consideration, then $X$ is called {\it non-singular}. 
If a $1$-form $\alpha$ does not vanish anywhere on some domain 
under consideration, then $\alpha$ is also called {\it non-singular}.  

\subsection{Riemannian and Lorentzian manifolds}
Riemannian and Lorentzian geometries provide 
a natural description of the covariant form of Maxwell's equations 
as written in various textbooks. 
This description is also employed throughout  this paper. 
In this subsection, Riemannian and Lorentzian geometries are 
summarized and the notations are fixed. 
These employed notations are based on several 
textbooks\,\cite{Nakahara,Frenkel}.
See also \cite{Lee2012,Choquet1982}.  

Let $(\cM,g)$ be a $4$-dimensional Lorentzian manifold with
$\cM= \cI\times\cN$ with $\cN$ being a $3$-dimensional manifold and
$\cI\subset\mbbR$, where $g$ denotes a Lorentzian metric on $\cM$. 
In addition, let $x=(x^{0},x^{1},x^{2},x^{3})$ be coordinates 
for $\cM$, and $c_{0}\in\mbbR$
a positive constant that physically expresses the speed of light in vacuo,
$t=x^{0}/c_{0}\in\mbbR$ physically express time, and
$$
g=-\dr x^{0}\otimes \dr x^{0}+\bm{g}.
$$
This $\bm{g}$ is assumed to satisfy 
$\cL_{\partial/\partial x^{0}}\bm{g}=0$ throughout this paper. At a fixed 
$x^{0}$, $\cN_{x^{0}}$ denotes the hypersurface  in $\cM$, and  
$(\cN_{x^{0}},\bm{g})$ becomes a Riemannian manifold. 
Physically, in this paper, 
$\cM$ and $\cN$ express spacetime and space, respectively.  
The metric $g$ can be generalized to
$g_{0}=\sum_{a,b=0}^{3}\eta_{\,ab}\theta^{a}\otimes \theta^{b}$, where
$\eta_{ab}=\text{diag}\{-1,1,1,1\}$ and $\{\theta^{0},\ldots,\theta^{3}\}$
is a coframe with $\cL_{\partial/\partial x^{0}}\theta^{a}=0$, $a=1,2,3$.  
As one of the choices of $\bm{g}$ for $\cN=\mbbR^{3}$, a standard one is 
\beq
\bm{g}_{\mbbR^{3}}
=\dr x^{1}\otimes \dr x^{1}
+\dr x^{2}\otimes \dr x^{2}
+\dr x^{3}\otimes \dr x^{3}.
\label{g-E^3}
\eeq
The inverse of $g_{0}$ is 
$g_{0}^{-1}=\sum_{a,b=0}^{3}\eta^{\,ab}X_{a}\otimes X_{b}$, where
$(\eta^{ab})$ is the inverse of $(\eta_{ab})$, and $\{X_{0},\ldots,X_{3}\}$
is the dual basis of $\{\theta^{0},\ldots,\theta^{3}\}$. 
The metric dual $\wt{\ }:\GTM\to\GamLamM{1}$ is introduced, and 
its inverse is also denoted $\wt{\ }:\GamLamM{1}\to\GTM$ by an abuse of 
notation:
$$
\wt{X}
=g(X,-),
\qquad
\wt{\alpha}
=g^{-1}(\alpha,-),
\qquad
\forall X\in \GTM,\quad
\forall \alpha\in \GamLamM{1}.
$$
Similarly, on $\cN$ the following are introduced:   
$$
\wt{X}^{\bm{g}}
=\bm{g}(X,-),\qquad
{\wt{\alpha}}^{\bm{g}}
=\bm{g}^{-1}(\alpha,-),\qquad
\forall X\in \GT{\cN},\
\forall \alpha \in \GamLam{\cN}{1}.
$$
These metric duals are isomorphisms, since $g$ and $\bm{g}$ are
non-degenerate. 
In addition, the Hodge maps associated with $g$ and $\bm{g}$ are denoted by
\beqa
\star
&:&\GamLamM{k}\to \GamLamM{4-k},\qquad
k=0,1,\ldots,4,
\non\\
\star_{3}
&:&
\GamLam{\cN}{k}\to \GamLam{\cN}{3-k},\qquad
k=0,1,2,3.
\non
\eeqa
There are several ways to define the Hodge map, and one of them is as follows. 
The action of $\star$  and that of $\star_{3}$  are determined by
\beqa
\star(\alpha_{1}\wedge\cdots\wedge\alpha_{r})
&=&\ii_{\wt{\alpha_{r}}}\cdots\ii_{\wt{\alpha_{1}}}\star 1,
\qquad 
\forall\ \alpha_{1},\cdots,\alpha_{r}\in\GamLamM{1}, 
\qquad
1\leq r\leq 4,
\non\\
\star_{3}(\beta_{1}\wedge\cdots\wedge\beta_{r})
&=&\ii_{\wt{\beta_{r}}}\cdots\ii_{\wt{\beta_{1}}}\star_{3} 1,
\qquad 
\forall\ \beta_{1},\cdots,\beta_{r}\in\GamLam{\cN}{1}, 
\qquad
1\leq r\leq 3,
\non
\eeqa
with fixed volume elements $\star 1$ and $\star_{3}1$. 
It then follows for  $\alpha\in\GamLamM{k}$, 
$(k=0,\ldots,3)$ and  $\beta\in\GamLam{\cN}{k}$, $(k=0,1,2)$  
that 
\beqa
\ii_{X}\star\alpha
&=&\star\left(\alpha\wedge\wt{X}\right),\qquad \forall\ X\in\GTM,
\non\\
\ii_{X}\star_{3}\beta
&=&\star_{3}\left(\beta\wedge\wt{X}^{\bm{g}}\right),\qquad \forall\ X\in\GT{\cN}.
\non
\eeqa
The volume elements are fixed as 
\beqa
\star 1
&=&\dr x^{0}\wedge\theta^{1}\wedge\theta^{2}\wedge\theta^{3}
=\dr x^{0}\wedge\star_{3}1,
\non\\
\star_{3}1
&=&\theta^{1}\wedge\theta^{2}\wedge\theta^{3}. 
\non
\eeqa
It can be shown from straightforward calculations  that
\beqa
&&\star_{3}\theta^{1} 
=\theta^{2}\wedge\theta^{3}, 
\qquad
\star_{3}\theta^{2} 
=\theta^{3}\wedge\theta^{1}, 
\qquad
\star_{3}\theta^{3} 
=\theta^{1}\wedge\theta^{2}, 
\non\\
&&\star_{3}\left(\theta^{2}\wedge\theta^{3}\right)
=\theta^{1}, 
\qquad
\star_{3}\left(\theta^{3}\wedge\theta^{1}\right)
=\theta^{2}, 
\qquad
\star_{3}\left(\theta^{1}\wedge\theta^{2}\right)
=\theta^{3}, 
\non
\eeqa
and
$$
\star_{3}
\left(\theta^{1}\wedge\theta^{2}\wedge\theta^{3}\right)
=1,
$$
from which $\star_{3}\star_{3}\beta=\beta$ for
all $\beta\in\GamLam{\cN}{k}$, ($k=0,\ldots,3$).
Similarly, it follows that 
$$
\star\star\alpha
=(-1)^{k+1}\alpha,\qquad\forall \alpha\in\GamLamM{k},\quad
k=0,\ldots,4.
$$
Note that, as a volume element,
the so-called {\it Riemannian volume form} 
$\star 1=\sqrt{|\det (g_{ab})|}\dr x^{0}\wedge\cdots\wedge\dr x^{3}$
is often chosen for $g=\sum_{a,b=0}^{3}g_{ab}\dr x^{a}\otimes\dr x^{b}$
with given $(g_{ab})$\,\cite{Lee2012}. 
The Riemannian volume form is sometimes simply called
the {\it volume form}\,\cite{Choquet1982}.

\subsection{Maxwell's equations}
\label{section-preliminaries-Maxwell-equation}
In physics, Maxwell's equations together with boundary conditions  
and given externally applied sources    
are employed to describe electromagnetic
fields\,\cite{Jackson1998,Landau1971}.
They are often written in the Gibbs notation,    
and their dynamical variables are of the form of vectors. 
These variables are so-called electric field written in this notation as
$\vec{E}=(E_{1},E_{2},E_{3})$,
magnetic field $\vec{H}=(H_{1},H_{2},H_{3})$, 
and other fields, where all the vector components are real-valued functions 
on space and time. The other fields are 
called magnetic flux $\vec{B}=(B_{1},B_{2},B_{3})$ 
and electric displacement field $\vec{D}=(D_{1},D_{2},D_{3})$, 
where $\vec{B}$ and $\vec{D}$ are related with $\vec{H}$ and $\vec{E}$. 
Such relations express electromagnetic properties of media, where
examples of media are dielectrics, conductors, and vacuum. In the case of
vacuum, the relations are given by $\vec{D}=\varepsilon_{0}\vec{E}$ and
$\vec{B}=\mu_{0}\vec{H}$ with $\varepsilon_{0}$ and $\mu_{0}$ being positive
constants. 
In the SI unit system, Maxwell's equations in vacuo for observers at rest 
are described by the 
partial differential equations with the Gibbs vector notation, 
$$
\vec{\nabla}\bm{\times}\vec{E}
=-\frac{\partial}{\partial t}\vec{B},\quad
\vec{\nabla}\bm{\cdot}\vec{B}
=0,\quad
\vec{\nabla}\bm{\cdot}\vec{D}
=0,\quad
\vec{\nabla}\bm{\times}\vec{H}
=\frac{\partial}{\partial t}\vec{D},
$$ 
where
$\vec{\nabla}=(\partial/\partial x,\partial/\partial y,\partial/\partial z)$, 
$\bm{\cdot}$ denotes the scalar product between two vectors, 
and $\bm{\times}$ denotes the outer product between two vectors
in $3$-dimensional manifolds. 
Meanwhile Maxwell's equations are written in terms of differential forms
on Lorentzian manifolds. 
Both of the descriptions have advantages, and the approach with
differential forms is well-fit to geometry for given
manifolds, and is employed in this paper. 
Let $(\cM,g)$ be a Lorentzian manifold, where
$\cM$ is of the form $\cM=\cI\times\cN$ and $g$ is of the form
$g=-\dr x^{0}\otimes \dr x^{0}+\bm{g}$. Here $(\cN,\bm{g})$ is 
a Riemannian manifold, and expresses physical space with a fixed time.
This $\cM$ models a spacetime physically. 
Since it is known that electromagnetic fields are invariant under
the Lorentzian transformation on spacetime, this Lorentzian manifold
is naturally fit to electromagnetism.  
As will be shown below, two $2$-forms $F_{0}$ and $F_{1}$
will be introduced on $\cM$. 
This $F_{0}$ provides information about $\vec{B}$ and $\vec{E}$,
and the other one, $F_{1}$, 
provides information  about $\vec{D}$ and $\vec{H}$. Here the roles of
$\vec{B}$ and $\vec{D}$
are played by $2$-forms $\bm{B}$ and $\bm{D}$, and the roles of
$\vec{E}$ and $\vec{H}$
are played by $1$-forms $\bm{e}$ and $\bm{h}$.

In what follows Maxwell's equations and their basic properties are 
summarized briefly in the language of differential forms.  
For the purpose of
studying Maxwell's equations in vacuo,
the following descriptions are somewhat redundant, however
this redundancy allows us to consider more general cases if needed,  
and this notation is  
commonly used in the literature\,\cite{Nakahara,Frenkel}.
Hence this commonly used notation is adopted in this paper.
Mathematical physics backgrounds
of physical quantities introduced in this subsection
can be found in\,\cite{Abraham1998,Benn1987,Heal2003,Gratus2017,Kitano2012,Burton2024}.

On a Lorentzian manifold $(\cM,g)$,
let $F_{0}$ and $F_{1}$ be $2$-forms on $\cM$, where they are related by
\beq
F_{1}
=-\varepsilon_{0}\star F_{0},
\label{constitutive-relation-vacuo-0}
\eeq
with $\varepsilon_{0}>0$ being the constant called 
the {\it electric permittivity} in vacuo. 
The $2$-form $F_{0}$ is called the {\it Maxwell $2$-form} and
$\star^{-1}F_{1}$ called the {\it excitation $2$-form}. 
In theoretical physics, when studying fields in vacuo, 
$F_{0}$ and $F_{1}$ are not distinguished, and they are denoted simply by $F$.
This $F$ is often called the {\it (electromagnetic) field strength}.
In addition $G:=\star^{-1}F_{1}$ is a commonly employed symbol in physics.
Relations between $F_{0}$ and $F_{1}$ or
ones between $F_{0}$ and $\star^{-1} F_{1}$ are called 
{\it (electromagnetic) constitutive relations}. 
For later use let $\mu_{0}>0$ be the constant called the 
{\it permeability} in vacuo, where this $\mu_{0}$ is not freely chosen.
The relation among $c_{0},\varepsilon_{0}$ and $\mu_{0}$ is
$$
c_{0}
=\frac{1}{\sqrt{\varepsilon_{0}\mu_{0}}}.
$$
For media that are not vacuum, 
constitutive relations between $F_{1}$ and $F_{0}$
are not \fr{constitutive-relation-vacuo-0}, 
and are chosen appropriately.  
Constitutive relations are often  
of the form $F_{1}=\star \kappa F_{0}$
with some $\kappa:\GamLamM{2}\to\GamLamM{2}$.
Note that 
$\kappa^{\,\prime}:=\star\kappa$ can also be introduced and employed, and 
in this case 
the constitutive relations are of the form 
$F_{1}=\kappa^{\,\prime}F_{0}$ with some  
$\kappa^{\,\prime}:\GamLamM{2}\to\GamLamM{2}$.
In addition, further generalizations of constitutive relations can be 
considered if needed.  

Suppose that $F_{0}$ and $F_{1}$ are decomposed as
\beqa
F_{0}&=&-c_{0}\bm{B}-\bm{e}\wedge\dr x^{0},
\label{F0}\\
F_{1}&=&\bm{D}-c_{0}^{-1}\bm{h}\wedge\dr x^{0},
\label{F1}
\eeqa
where  
$\bm{B},\bm{D}\in\GamLamM{2}$ and $\bm{e},\bm{h}\in\GamLamM{1}$  
satisfy
$$
\ii_{\partial/\partial x^{0}}\bm{B}
=\ii_{\partial/\partial x^{0}}\bm{D}
=\ii_{\partial/\partial x^{0}}\bm{e}
=\ii_{\partial/\partial x^{0}}\bm{h}
=0.
$$
It then follows that one can extract $\bm{e}$ and $\bm{h}$ from $F_{0}$ and
$F_{1}$ as 
$$
\bm{e}
=\ii_{\partial/\partial x^{0}}F_{0},\qquad\text{and}\qquad
\bm{h}
=c_{0}\ \ii_{\partial/\partial x^{0}}F_{1}.
$$
In physics, these $1$-form and $2$-form fields are termed 
as shown in Table\,\ref{table-forms-names}. In addition, roughly speaking, 
the $1$-forms $\bm{e}$ and $\bm{h}$ determine polarization
if the corresponding electromagnetic fields are traveling waves.  
\begin{table}[htb]
\caption{Differential forms for Maxwell's equations}
\label{table-forms-names}
  \begin{center}
    \begin{tabular}{|c|c|}
      \hline
      $F_{0}\in\GamLamM{2}$& Maxwell $2$-form\\
      \hline
      $\star^{-1}F_{1}\in\GamLamM{2}$& Excitation $2$-form\\
      \hline
      $j\in\GamLamM{3}$& Source $3$-form\\
      \hline
      $\bm{B}\in\GamLamM{2}$& Magnetic flux field\\
      \hline
      $\bm{D}\in\GamLamM{2}$& Electric displacement field\\
      \hline
      $\bm{e}\in\GamLamM{1}$& Electric field\\
      \hline
      $\bm{h}\in\GamLamM{1}$& Magnetic field\\
      \hline
    \end{tabular}
  \end{center}    
\end{table}  

Throughout this paper we focus on the following:
\begin{Definition}
\label{definition-Maxwell's-equations}
  {\it Maxwell's equations
    in vacuo on a Riemannian manifold $(\cM,g)$} are
\beq
\dr F_{0}
=0,\qquad\text{and}\qquad
\dr F_{1}
=0.
\label{Maxwell-equations-vacuo}
\eeq
Solutions $F_{0}$ and $F_{1}$, or a set of fields
$\bm{e},\bm{h},\bm{D},\bm{B}$, are called {\it Maxwell fields}.
With respect to the coordinate system $x=(x^{0},\ldots,x^{3})$, 
  Maxwell fields are called {\it time-independent} if
  $\cL_{\partial/\partial x^{0}}F_{0}=0$ and $\cL_{\partial/\partial x^{0}}F_{1}=0$.
  Otherwise Maxwell fields are called {\it time-dependent}. 
\end{Definition} 
In Definition\,\ref{definition-Maxwell's-equations}, 
although $F_{0}$ and $F_{1}$ seem mathematically similar,  
a dissimilarity between $F_{0}$ and $F_{1}$ is revealed when 
one distinguishes pseudoness associated with differential 
forms\,\cite{Kitano2012,Nakata2019,Gratus2017}.    
From this distinction, forms are classified into two types,  
one class is the set consisting of untwisted forms 
and the other class is that of twisted forms.  
As discussed in\,\cite{Kitano2012,Nakata2019}, 
$F_{0}$ is untwisted, and $F_{1}$ is twisted. 
Then for a given medium, a relation is introduced to link
the untwisted form $F_{0}$ and  
the twisted form $F_{1}$, and the Hodge map flips their pseudoness.  
In the case of vacuum, this relation 
can be written in terms of the so-called 
vacuum admittance or the vacuum impedance. 
In this paper 
forms are not classified according to pseudoness for brevity, 
and this classification is not emphasized.  
Rather than this classification, we will 
concentrate attention on the symplectic and its related properties.   

For an electromagnetic system with an external source, Maxwell's equations are
$\dr F_{0}=0$ and $\dr F_{1}=j$ with $j$ being a $3$-form that physically
expresses an external  source. In this paper $j=0$ everywhere on $\cM$ is 
assumed throughout. 

To rewrite Maxwell's equations as a decomposed form, introduce
$$
\bm{\dr}:\GamLam{\cM}{k}\to
\GamLam{\cM}{k+1},\qquad k=0,1,\ldots,4,
$$
such that 
$$
\dr=
\bm{\dr}+\dr x^{0}\wedge \cL_{\partial/\partial x^{0}}.
$$
For example, if $f$ is a function of $x=(x^{0},\ldots,x^{3})$, then
$$
{\bm\dr} f
=\sum_{a=1}^{3}\frac{\partial f}{\partial x^{a}}\dr x^{a}.
$$
By an abuse of notation, the operator $\bm{\dr}$ is also used for
$k$-forms on $3$-dimensional manifolds $\cN$, that is,
$\bm{\dr}:\GamLam{\cN}{k}\to\GamLam{\cN}{k+1}$, $k=0,1,2,3$. 

Then Maxwell's equations \fr{Maxwell-equations-vacuo} with
\fr{F0} and \fr{F1} are decomposed into 
\beq
\bm{\dr}\bm{e}
=-\cL_{\partial/\partial t}\bm{B},\qquad
\bm{\dr}\bm{B}
=0,\qquad
\bm{\dr}\bm{D}
=0,\qquad
\bm{\dr}\bm{h}
=\cL_{\partial/\partial t}\bm{D},
\label{Maxwell-decomposed-vacuo}
\eeq
(see Section\,\ref{section-appendix-Maxwell-decomposed} 
for their derivations).  
In vacuo, the constitutive relation $F_{1}=\star\kappa F_{0}$ 
is implemented as \fr{constitutive-relation-vacuo-0} 
by $\kappa=-\varepsilon_{0}\Id$, where 
$\Id:\GamLamM{2}\to\GamLamM{2}$ is the identity operator.
The constitutive relation for vacuo leads to  
\beq
\bm{D}=\varepsilon_{0}\star_{3}\bm{e},\qquad\text{and}\qquad
\bm{B}=\mu_{0}\star_{3}\bm{h},
\label{Maxwell-constitutive-relations-vacuo-decomposed}
\eeq
(see Section\ref{section-appendix-constitutive-relations-decomposed} 
for their derivations).  
For Maxwell fields, energy density forms are one of important quantities.
In the case of vacuo, they are $3$-forms defined as 
\beq
\bm{\cE}_{e}
:=\frac{\varepsilon_{0}}{2}\bm{e}\wedge\star_{3}\bm{e},\qquad
\bm{\cE}_{h}
:=\frac{\mu_{0}}{2}\bm{h}\wedge\star_{3}\bm{h},
\label{energy-densities}
\eeq
which can be generalized to systems in some media as 
\beq
\bm{\cE}_{e}^{\kappa}
:=\frac{1}{2}\bm{e}\wedge\bm{D},\qquad
\bm{\cE}_{h}^{\kappa}
:=\frac{1}{2}\bm{h}\wedge\bm{B}.
\label{energy-densities-general-1}
\eeq
In addition, the $2$-form defined as 
\beq
\bm{\cS}  
:=\bm{e}\wedge\bm{h},
\label{Poynting-2-form}
\eeq
is called the {\it Poynting form} 
in this paper, and this
corresponds to the Poynting vector in electromagnetism.   

The following is an example of a Maxwell system in vacuo.
\begin{Example}
\label{Example-traveling-solution}
Let $f$ be a function on $\cM=\mbbR^{4}$, and 
\beqa
\bm{B}
&=&c_{0}^{-1}f(x^{3}-x^{0})\,\dr x^{3}\wedge\dr x^{1},
\non\\
\bm{D}
&=&\varepsilon_{0}f(x^{3}-x^{0})\,\dr x^{2}\wedge\dr x^{3},
\non\\
\bm{e}
&=&f(x^{3}-x^{0})\,\dr x^{1},
\non\\
\bm{h}
&=&\frac{1}{c_{0}\mu_{0}}f(x^{3}-x^{0})\,\dr x^{2}.
\non
\eeqa
Straightforward calculations show that these fields satisfy the relations
$$
\bm{B}
=\mu_{0}\star_{3}\bm{h},\qquad\text{and}\qquad
\bm{D}
=\varepsilon_{0}\star_{3}\bm{e}.
$$
Then $F_{0}$ and $F_{1}$ constructed from these forms
\beqa
F_{0}
&=&f(x^{3}-x^{0})\dr x^{1}\wedge \dr(x^{3}-x^{0}),
\non\\
F_{1}
&=&\varepsilon_{0} f(x^{3}-x^{0})\dr x^{2}\wedge \dr(x^{3}-x^{0}),
\non
\eeqa
are Maxwell, due to $\dr F_{0}=0$ and $\dr F_{1}=0$. 
If $f(\xi)=\sin(\xi)$, then these fields express
traveling waves. Note that
\beq
F_{0}\wedge F_{0}
=0,\qquad\text{and}\qquad
F_{1}\wedge F_{1}
=0.
\label{Maxwell-traveling-F-non-symplectic}
\eeq
\end{Example}

Recall that symplectic forms are defined on even dimensional manifolds
such that  
these forms are $2$-forms that are closed and non-degenerate,  where 
non-degenerate conditions for $2$-forms $F_{0}$ and $F_{1}$ on a
$4$-dimensional manifold are equivalent to $F_{0}\wedge F_{0}\neq 0$ and
$F_{1}\wedge F_{1}\neq 0$ at any point. 
In many cases, including Example\,\ref{Example-traveling-solution}, 
$2$-forms $F_{0}$ and $F_{1}$ are not symplectic, 
(see \fr{Maxwell-traveling-F-non-symplectic}). 
Meanwhile some Maxwell fields are symplectic. 
In such a case symplectic topological methods are expected to 
reveal topological aspects of Maxwell fields.
Similarly, if Maxwell fields have another geometric structure,
then its corresponding topology can be employed. 
In Section\,\ref{section-symplectic-contact-topology},
some basics of contact and symplectic topology
is summarized for discussing Maxwell fields.  

\subsection{Symplectic and contact topology}
\label{section-symplectic-contact-topology}
In this paper
symplectic topology is considered on $4$-dimensional manifolds,
and contact one is
considered on $3$-dimensional manifolds.
Some of discussions in this section 
can be generalized to even and odd dimensional cases\,\cite{McDuff}. 

The following are standard in symplectic and contact geometries.
\begin{Definition}
  Let $\cM$ be a $4$-dimensional manifold, and $\omega$ a $2$-form
  on $\cM$. If $\omega$ satisfies $\omega\wedge\omega\neq 0$ anywhere, then
  $\omega$ is called {\it non-degenerate}.
  If $\omega$ is closed and non-degenerate, equipped on $\cM$,  then the pair
  $(\cM,\omega)$ is called a {\it symplectic manifold}.
\end{Definition}
\begin{Definition}
  Let $\cN$ be a $3$-dimensional manifold, and $\lambda$ a $1$-form on $\cN$.
  If $\lambda$ satisfies $\lambda\wedge\bm{\dr}\lambda\neq 0$ anywhere
  on $\cN$, 
  then $\lambda$ is called a {\it contact form}, and
  $\ker\lambda:=\{X\in T\cN\,|\, \lambda(X)=0\}$ is called 
  a {\it contact structure}. The pair $(\cN,\lambda)$ or $(\cN,\ker\lambda)$
  is called a {\it contact manifold}. If $\lambda$ is globally defined, then
  it is called {\it co-oriented}.
\end{Definition}
Given a contact form $\lambda$ on $\cN$, there exists a unique vector field $Y$
such that
$$
\ii_{Y}\bm{\dr}\lambda
=0,\qquad\text{and}\qquad
\ii_{Y}\lambda
=1.
$$
This vector field, $Y$, is called the {\it Reeb vector field}. 
The normalization condition $\ii_{Y}\lambda=1$ can be relaxed as follows. 
If $Y$ is such that
$$
\ii_{Y}\bm{\dr}\lambda
=0,\qquad \text{and}\qquad
\ii_{Y}\lambda
>0,
$$
then this $Y$ is called a {\it Reeb-like vector field} \cite{Etnyre2000}. 

One of the most basic questions in dynamical systems theory is to find 
a closed periodic orbit in phase space for a system under consideration.  
The Weinstein conjecture is a conjecture regarding this kind of a question
for Reeb dynamics on contact manifolds
(see \cite{Weinstein1979} for the original version). 
\begin{quote}
{\bf The Weinstein conjecture:} 
Let $(\cN,\lambda)$ be a closed co-oriented contact manifold.
Then the Reeb vector field admits a closed periodic orbit.
\end{quote}

In symplectic manifolds, classes of hypersurfaces are often discussed.
One of them is the class of contact type hypersurfaces. 
A useful generalization of the notion of 
a hypersurface of contact type is known as
follows\,\cite{McDuff,Cieliebak2015}. 

\begin{Definition}
\label{definition-stable-Hamiltonian-structure} 
A {\it stable Hamiltonian structure} on a manifold $\cN$ of dimension $3$ 
is a pair $(\Omega,\lambda)$, consisting of a closed $2$-form $\Omega$ 
and a $1$-form $\lambda$, such that 
$$ 
\lambda\wedge \Omega
\neq 0,\qquad
\ker\Omega\subset \ker{\bm\dr}\lambda.  
$$
The corresponding Reeb vector field is the unique
vector field $Y$ on $\cN$ such that
$$
\ii_{Y}\Omega
=0,\qquad\text{and}\qquad
\ii_{Y}\lambda
=1.  
$$
\end{Definition}
  To show examples of $(\Omega,\lambda)$
  in Definition\,\ref{definition-stable-Hamiltonian-structure},  
  several preliminary statements are needed. Hence, 
  explicit realizations of stable Hamiltonian structures are not shown here. 
  They will be shown in Section\,\ref{section-example-Beltrami-form}. 

The following remark is about
Definition\,\ref{definition-stable-Hamiltonian-structure}.  
\begin{Remark}
\label{remark-stable-Hamiltonian-structure}
1. A manifold is assumed to be closed sometimes in the literature.
2. The condition $\ker\Omega\subset \ker\bm{\dr}\lambda$ is written as 
$\bm{\dr}\lambda=f^{\Omega}\Omega$ 
with $f^{\Omega}$ being some function. 
This implies that $\ii_{Y}\bm{\dr}\lambda=0$ and $\ii_{Y}\Omega=0$.
From these together with   
$\ii_{Y}\lambda=1$ and $\bm{\dr}\bm{\Omega}=0$, it follows that 
$$
\cL_{Y}\lambda
=(\bm{\dr}\ii_{Y}+\ii_{Y}\bm{\dr})\lambda
=0,\quad\text{and}\quad
\cL_{Y}\Omega
=(\bm{\dr}\ii_{Y}+\ii_{Y}\bm{\dr})\Omega
=0.
$$
Hence, 
$$
\cL_{Y}(\lambda\wedge\Omega)
=(\cL_{Y}\lambda)\wedge\Omega
+\lambda\wedge(\cL_{Y}\Omega)
=0.
$$
3. A contact form $\lambda$ determines 
a stable Hamiltonian structure
by choosing $\Omega=\bm{\dr}\lambda$ with $f^{\Omega}=1$ at every point of 
$\cN$.
4. A contact form is a special case of a stable Hamiltonian structure. 
\end{Remark}

Integral curves of the Reeb vector fields on manifolds with
stable Hamiltonian structures are considered in this paper.  
To see these curves, one expresses  the Reeb vector field  
$Y$ in terms of
$\Omega$ and $\lambda$ in coordinates.  
Write $\Omega$, $\lambda$, and $Y$ in coordinates as 
\beqa
\lambda
&=&\lambda_{1}\dr x^{1}+\lambda_{2}\dr x^{2}+\lambda_{3}\dr x^{3},
\non\\
\Omega
&=&
\Omega_{3}\dr x^{1}\wedge\dr x^{2}
+\Omega_{1}\dr x^{2}\wedge\dr x^{3}
+\Omega_{2}\dr x^{3}\wedge\dr x^{1}, 
\non
\eeqa
and
$$
Y=
Y^{1}\frac{\partial}{\partial x^{1}}+
Y^{2}\frac{\partial}{\partial x^{2}}+
Y^{3}\frac{\partial}{\partial x^{3}},
$$
where $\lambda_{a},\Omega_{a},Y^{a}$, ($a=1,2,3$) are functions on $\cN$. 
The condition $\lambda\wedge\Omega\neq 0$ yields   
$$
\lambda_{1}\Omega_{1}+\lambda_{2}\Omega_{2}+\lambda_{3}\Omega_{3}
\neq 0.
$$
In addition, the  
conditions for $Y$ yield
$$
\lambda_{1}Y^{1}+\lambda_{2}Y^{2}+\lambda_{3}Y^{3}
=1,
$$
and
$$
\Omega_{2}Y^{3}-\Omega_{3}Y^{2}
=0,\quad
\Omega_{3}Y^{1}-\Omega_{1}Y^{3}
=0,\quad
\Omega_{1}Y^{2}-\Omega_{2}Y^{1}
=0.
$$
One can determine $Y^{a}$ by solving these equations for $Y^{a}$. 
For instance, when $\Omega_{3}\neq 0$,  
one has that
\beq
Y^{1}
=\frac{\Omega_{1}}{\Omega_{3}}Y_{3},\qquad
Y^{2}
=\frac{\Omega_{2}}{\Omega_{3}}Y_{3},\qquad
Y^{3}
=\frac{\Omega_{3}}{\lambda_{1}\Omega_{1}+\lambda_{2}\Omega_{2}+\lambda_{3}\Omega_{3}}.
\label{Reeb-component-SHS-omega-3-not-0}
\eeq
When $\Omega_{1}\neq0$, $\Omega_{2}\neq0$, and $\Omega_{3}=0$,  
one has that
$$
\lambda_{1}\Omega_{1}+\lambda_{2}\Omega_{2}
\neq 0,\quad
\Omega_{2}Y^{3}
=0,\quad\text{and}\quad
\Omega_{1}Y^{3}
=0,
$$
from which $Y^{3}=0$. Solving the set of the equations given by  
$$
\lambda_{1}Y^{1}+\lambda_{2}Y^{2}
=1,\quad\text{and}\quad 
\Omega_{2}Y^{1}-\Omega_{1}Y^{2}
=0,
$$
one has that 
\beq
Y^{1}
=\frac{\Omega_{1}}{\lambda_{1}\Omega_{1}+\lambda_{2}\Omega_{2}},\qquad
Y^{2}
=\frac{\Omega_{2}}{\lambda_{1}\Omega_{1}+\lambda_{2}\Omega_{2}},
\qquad
Y^{3}=0.
\label{Reeb-component-SHS-omega-3=0}
\eeq
Second, identify $Y^{a}$ as 
$$
Y^{a}=\frac{\dr x^{a}}{\dr s},\qquad a=1,2,3,
$$
with $s\in\mbbR$. Then solutions to the system of
ordinary differential equations (ODEs) given by  
\fr{Reeb-component-SHS-omega-3-not-0},
$$
\frac{\dr x^{1}}{\dr s}
=\frac{\Omega_{1}}{\sum_{a=1}^{3}\lambda_{a}\Omega_{a}},
\quad
\frac{\dr x^{2}}{\dr s}
=\frac{\Omega_{2}}{\sum_{a=1}^{3}\lambda_{a}\Omega_{a}},
\quad
\frac{\dr x^{3}}{\dr s}
=\frac{\Omega_{3}}{\sum_{a=1}^{3}\lambda_{a}\Omega_{a}},
$$
and solutions to that given by \fr{Reeb-component-SHS-omega-3=0},
$$
\frac{\dr x^{1}}{\dr s}
=\frac{\Omega_{1}}{\sum_{a=1}^{2}\lambda_{a}\Omega_{a}},
\quad
\frac{\dr x^{2}}{\dr s}
=\frac{\Omega_{2}}{\sum_{a=1}^{2}\lambda_{a}\Omega_{a}},
\quad
\frac{\dr x^{3}}{\dr s}
=0,
$$
express orbits of the Reeb vector field $Y$.
The Weinstein conjecture and its related theorems state that
there are some periodic orbits in the dynamical systems. 

Taubes showed the following:
\begin{Theorem}[Taubes\,\cite{Taubes2007}]  
\label{fact-Taubes-2007}
Let $(\cN,\lambda)$ be a compact $3$-dimensional contact manifold with
$\lambda$ being a contact form.
Then the Reeb vector field has a closed integral curve.  
\end{Theorem}
Since the proof by Taubes consists of a large body, 
the survey paper \cite{Hutchings2010} by Hutchings
helps prepare the reader to learn the full story. 
Hutchings and Taubes \cite{Hutchings2009} then showed the following:
\begin{Theorem}[Theorem 1.1 of \cite{Hutchings2009}] 
\label{fact-Hutchings2009} 
Let $\cN$ be a closed oriented connected $3$-manifold with a stable 
Hamiltonian structure. If $\cN$ is not a $T^{2}$-bundle over $S^{1}$, 
then the associated Reeb vector field has a closed orbit. 
\end{Theorem}

Theorems\,\ref{fact-Taubes-2007} and \ref{fact-Hutchings2009}
will be employed in proving the existence of a
closed integral curve of the vector field
$\wt{\bm{e}}^{\bm{g}}$ and that of $\wt{\bm{h}}^{\bm{g}}$  
at a fixed time,
where these vector fields will be identified with Reeb vector fields. 

In general, the Maxwell fields depend on time, and they are described 
on $4$-dimensional manifolds. For this reason 
the study on a $3$-dimensional manifold $\cN$ discussed so far 
is not general enough.   
Therefore it is desirable to describe
Maxwell fields on $4$-dimensional manifolds,
and hence relations between $4$-dimensional
symplectic manifolds and Maxwell fields will be discussed 
in Section\,\ref{section-symplectic-SHS}.  

\subsection{Geometry related to Beltrami fields}
\label{section-geometry-Beltrami}
In the following, the necessary background on 
Beltrami fields on $3$-dimensional Riemannian manifolds is 
summarized.

The following vector filed and its metric dual underpin  
discussions on geometry of Maxwell fields in this paper.  

\begin{Definition}
\label{definition-Beltrami-1-form-f}
Let $(\cN,\bm{g})$ be a $3$-dimensional Riemannian manifold.  
A $1$-form $\bm{v}$ on $\cN$ is a {\it Beltrami $1$-form} if 
\beq
\star_{3}\bm{\dr}{\bm v}
=f\bm{v},
\label{Beltrami-1-form-general}
\eeq
with some $f\in\GamLam{\cN}{0}$.
A {\it rotational Beltrami $1$-form} is one for which
$$
f\neq 0.
$$
A {\it Beltrami vector field} is the metric dual of a Beltrami $1$-form,
and a {\it rotational Beltrami field} is for which $f\neq 0$. 
\end{Definition}
In some references the Beltrami $1$-form 
is defined as a $1$-form on $3$-dimensional Riemannian manifolds  
that satisfies  \fr{Beltrami-1-form-general} and
\beq
\star_{3}\bm{\dr}\star_{3}\bm{v}
=0,
\label{Beltrami-1-form-general-divergence-free}
\eeq
(see \cite{Slobodeanu2023}).
Suppose that $\bm{v}$ is a rotational Beltrami $1$-form with $f$ 
as in Definition\,\ref{definition-Beltrami-1-form-f}. In addition 
suppose that this $f$ is non-zero constant, and then in this case, 
\fr{Beltrami-1-form-general-divergence-free} is satisfied, due to 
$$
\bm{\dr}\star_{3}\bm{v}
=\bm{\dr}\star_{3}\left(\frac{1}{f}\star_{3}\bm{\dr}\bm{v}\right)
=\frac{1}{f}\bm{\dr}\star_{3}\star_{3}\bm{\dr}\bm{v}
=0.
$$
Hence a rotational Beltrami $1$-form with constant $f\neq 0$
satisfies \fr{Beltrami-1-form-general-divergence-free}.

As is well-known \cite{Martinet1971}, every
oriented compact smooth $3$-manifold admits a contact structure. 
Then in \cite{Etnyre2000}, 
relations between Beltrami-like fields and
contact $3$-manifolds have been shown. 
\begin{Theorem}[Theorem 2.1 of \cite{Etnyre2000}]
\label{fact-Etnyre2000}
Let $(\cN,\bm{g})$ be a Riemannian $3$-manifold. 
Any smooth, non-singular rotational Beltrami vector field on $\cN$ 
is a Reeb-like vector field for some contact form on $\cN$.  
Conversely, given a contact form $\lambda$ on $\cN$ with Reeb vector field
$Y$, any non-zero rescaling of $Y$ is a smooth, non-singular  rotational  
Beltrami vector field for some Riemannian metric on $\cN$. 
\end{Theorem}

Note that in Theorem\,\ref{fact-Etnyre2000},
on a given $(\cN,\bm{g})$ a non-singular rotational Beltrami
vector field is assumed to exist. Meanwhile it is unclear whether there
exists this vector field on $\cN$.  
If $f$ in \fr{Beltrami-1-form-general}
is constant, then the Beltrami $1$-form 
is recognized as an eigen $1$-form associated with the operator 
$\star_{3}\bm{\dr}:\GamLam{\cN}{1}\to\GamLam{\cN}{1}$.
Therefore, in the case that $f$ is constant, 
one should solve the eigenvalue problem on the manifold $\cN$
to see an explicit form of $\bm{v}$  
\cite{Enciso2012}. If $f$ is a non-trivial function, not a constant, then  
more discussions are needed  \cite{Enciso2016}.

The following shows how to construct a contact form
from a special non-singular rotational Beltrami $1$-form.
\begin{Proposition}[\cite{Etnyre2000}]
Let $(\cN,\bm{g})$ be a $3$-dimensional Riemannian manifold, and 
$\bm{v}$ a non-singular $1$-form that satisfies 
$$
\star_{3}\bm{\dr}\bm{v}
=k\bm{v},
$$
with $k\neq0$ being constant.  
Then $\bm{v}$ is a contact form on $\cN$. 
\end{Proposition}
\begin{Proof}
  Since $\bm{v}=k^{-1}\star_{3}\bm{\dr}\bm{v}$, one has that 
  $$
  0\neq
  \bm{g}^{-1}(\bm{v},\bm{v})\star_{3}1
  =\bm{v}\wedge\star_{3}\bm{v}
  =\bm{v}\wedge\star_{3}(k^{-1}\star_{3}\bm{\dr}\bm{v})
  =k^{-1}\bm{v}\wedge\bm{\dr}\bm{v}.
  $$
  This shows that $\bm{v}\wedge\bm{\dr}\bm{v}\neq 0$, from which
  $\bm{v}$ is a contact form. 
  \qed
\end{Proof}

The following shows how to construct a stable Hamiltonian structure 
from a special non-singular rotational Beltrami $1$-form.
\begin{Proposition}
\label{fact-stable-Hamiltonian-structure-by-Beltrami}
Let $(\cN,\bm{g})$ be a $3$-dimensional Riemannian manifold, and
$\bm{v}$ a non-singular $1$-form that satisfies
$$
\star_{3}\bm{\dr}\bm{v}
=k\bm{v},
$$
with $k\neq0$ being constant.  
Then $(\star_{3}\bm{v},\bm{v})$ is a stable Hamiltonian structure. 
In addition, let $Y$ be the Reeb vector field, and $Z$ the 
Reeb-like vector field such that 
$\ii_{Z}\bm{v}=\bm{g}^{-1}(\bm{v},\bm{v})$ and $\ii_{Z}\star_{3}\bm{v}=0$. 
It then follows that $Y=Z/\bm{g}^{-1}(\bm{v},\bm{v})$ 
with $Z={\wt{\bm{v}}}^{\bm{g}}$. 
\end{Proposition}
\begin{Proof}
First, we show that the pair $(\star_{3}\bm{v},\bm{v})$ satisfies 
the conditions for a stable Hamiltonian structure. They are   
(i) $\bm{\dr}\star_{3}\bm{v}=0$, (ii) $(\star_{3}\bm{v})\wedge\bm{v}\neq 0$,
and (iii) there is a function 
$f^{(0)}$ such that $\star_{3}\bm{v}=f^{(0)}\bm{\dr}\bm{v}$.

(Proof for i). 
It follows that
$$
\bm{\dr}\star_{3}\bm{v}
=\bm{\dr}\star_{3}\left(k^{-1}\star_{3}\bm{\dr}\bm{v}  \right)
=0.
$$

(Proof for ii).   
It follows that
$$ 
(\star_{3}\bm{v})\wedge\bm{v} 
=\bm{v}\wedge  \star_{3}\bm{v} 
=\bm{g}^{-1}(\bm{v},\bm{v})\,\star_{3}1 
\neq 0.  
$$ 
  
(Proof for iii).  
It follows that
$$
\star_{3}\bm{v}
=\star_{3}(k^{-1}\star_{3}\bm{\dr}\bm{v})
=k^{-1}\bm{\dr}\bm{v}.
$$
Hence $f^{(0)}$ is found to be $k^{-1}\neq 0$.
Second, the Reeb vector
field is shown to be of the form $Y=Z/\bm{g}^{-1}(\bm{v},\bm{v})$. 
Let $Z$ be a vector field satisfying $\ii_{Z}\bm{v}=\bm{g}^{-1}(\bm{v},\bm{v})$
and $\ii_{Z}\star_{3}\bm{v}=0$. 
To show the relation $Z={\wt{\bm{v}}}^{\bm{g}}$,
observe first that the equality holds:
$$
(\star_{3}\bm{v})\wedge\bm{v} 
=\bm{g}^{-1}(\bm{v},\bm{v})\star_{3}1. 
$$
Then, acting $\ii_{Z}$ on the left and right hand sides, one has that 
$$
\ii_{Z}[(\star_{3}\bm{v})\wedge\bm{v}] 
=(\star_{3}\bm{v})\ii_{Z}\bm{v} 
=\bm{g}^{-1}(\bm{v},\bm{v})\,\star_{3}\bm{v},  
$$
and that  
$$
\ii_{Z}[  \bm{g}^{-1}(\bm{v},\bm{v})\star_{3}1]  
=\bm{g}^{-1}(\bm{v},\bm{v})\star_{3}{\wt{Z}}^{\bm{g}}.
$$
By comparing the both sides, one has that 
$$
\star_{3}\bm{v}
=\star_{3}{\wt{Z}}^{\bm{g}},
$$
from which
$$
Z
={\wt{\bm{v}}}^{\bm{g}}.
$$
By comparing the definitions of $Y$ and $Z$
and the uniqueness of the Reeb vector field,  
$$
\ii_{Y}\bm{v}
=1,
\qquad\text{and}\qquad
\frac{1}{\bm{g}^{-1}(\bm{v},\bm{v})}\ii_{Z}\bm{v}
=1, 
$$
one immediately has
$Y=Z/\bm{g}^{-1}(\bm{v},\bm{v})$. 
\qed
\end{Proof}

The Reeb vector field $Y$ 
in Proposition\,\ref{fact-stable-Hamiltonian-structure-by-Beltrami},
where $(\Omega,\lambda)$ is a stable Hamiltonian structure with
$\lambda=\bm{v}$ and $\Omega=\star_{3}\bm{v}$, 
is written in coordinates as follows. 
Write $\Omega,\bm{v}$ and $Y$ as 
$$
\Omega
=\Omega_{3}\dr x^{1}\wedge\dr x^{2}
+\Omega_{1}\dr x^{2}\wedge\dr x^{3}
+\Omega_{2}\dr x^{3}\wedge\dr x^{1},
\qquad
\bm{v}
=v_{1}\dr x^{1}+v_{2}\dr x^{2}+v_{3}\dr x^{3},
$$
and 
$$
Y=Y^{1}\frac{\partial}{\partial x^{1}}+
Y^{2}\frac{\partial}{\partial x^{2}}+
Y^{3}\frac{\partial}{\partial x^{3}},  
$$
with $\Omega_{a},v_{a},Y^{a}\in\GamLam{\cN}{0}$, ($a=1,2,3$). 
Then it follows from $\Omega=\star_{3}\bm{v}$ that $\Omega_{a}=v_{a}$ for
all $a$. Hence $Y^{a}$ can be determined. For instance if $v_{3}=0$
and $\sum_{a=1}^{2}\lambda_{a}\Omega_{a}\neq 0$, then
$Y^{1}$ and $Y^{2}$ are given by 
\fr{Reeb-component-SHS-omega-3=0} with $\Omega_{1}=v_{1}$ and $\Omega_{2}=v_{2}$.

From Proposition\,\ref{fact-stable-Hamiltonian-structure-by-Beltrami},  
one has the following.
\begin{Corollary}
\label{fact-stable-Hamiltonian-structure-by-Beltrami-additional}  
Let $(\cN,\bm{g})$ be a $3$-dimensional Riemannian manifold,
and $\bm{v}$ a non-singular $1$-form that satisfies
$\star_{3}\bm{\dr}\bm{v}=k\bm{v}$ 
with $k$ being non-zero constant. In addition let $f$ be a non-vanishing
function on $\cN$. 
Then
$(\star_{3}\bm{v},f\bm{v})$ is a stable Hamiltonian structure, and the Reeb
vector field $Y^{f}$ is 
$$ 
Y^{f} 
=\frac{\wt{\bm{v}}^{\bm{g}}}{f\bm{g}^{-1}(\bm{v},\bm{v})}. 
$$ 
In particular, 
for the stable Hamiltonian structure 
$(\star_{3}\bm{v},\bm{v}/\bm{g}^{-1}(\bm{v},\bm{v}))$,
the Reeb vector field $Z$ is 
$$
Z=\wt{\bm{v}}^{\bm{g}}.
$$
\end{Corollary}
\begin{Proof}
The statement for $Y^{f}$ 
can be verified by substituting the expression of $Y^{f}$ into 
the conditions that the Reeb vector field should satisfy. 
Then choosing $f=1/\bm{g}^{-1}(\bm{v},\bm{v})$, one verifies that 
$Z=\wt{\bm{g}}^{\bm{g}}$ is the Reeb vector field. 
\qed
\end{Proof}

\subsection{Examples of Beltrami $1$-form}
\label{section-example-Beltrami-form}

The following are examples of 
Proposition\,\ref{fact-stable-Hamiltonian-structure-by-Beltrami}
and Corollary\,\ref{fact-stable-Hamiltonian-structure-by-Beltrami-additional}.
\begin{Example}
Consider the $3$-torus $\cN=T^{3}=S^{1}\times S^{1}\times S^{1}$ and the
metric 
$$
\bm{g}
=\dr x^{1}\otimes\dr x^{1}+\dr x^{2}\otimes\dr x^{2}+\dr x^{3}\otimes\dr x^{3},
$$
on $T^{3}$, that is the induced metric tensor field from the standard
one on $\mbbR^{3}$. 
Let $\bm{v}_{n}$ be the $1$-form with a fixed $n\in\mbbZ_{>0}$ 
$$
\bm{v}_{n}
=c(\,\cos(nx^{3})\dr x^{1}+\sin(nx^{3})\dr x^{2}),
$$
with $c>0$ being constant.
As has been stated in  \cite{Etnyre2000}, this $\bm{v}_{n}$
can be employed for the most general case of Beltrami flows on $T^{3}$
due to Theorem 4.1 of \cite{Etnyre2000}, where this theorem has been proved
by Giroux\,\cite{Giroux1994} and Kanda\,\cite{Kanda1997}.  
This $\bm{v}_{n}$ has the property that 
$$
\bm{g}^{-1}(\bm{v}_{n},\bm{v}_{n})
=c^{2}.
$$
Then the $1$-form $\bm{v}_{n}$ satisfies
$$
\star_{3}\bm{\dr}\bm{v}_{n}
=-n\bm{v}_{n}, 
$$
due to 
\beqa
\bm{\dr}\bm{v}_{n}
&=&-cn(\,\sin(nx^{3})\dr x^{3}\wedge\dr x^{1}
+\cos(nx^{3})\dr x^{2}\wedge \dr x^{3}),
\non\\
\star_{3}\bm{\dr}\bm{v}_{n}
&=&-cn(\,\sin(nx^{3})\dr x^{2}+\cos(nx^{3})\dr x^{1})
=-n\bm{v}_{n}.
\non
\eeqa
This shows that $k$ in 
Proposition\,\ref{fact-stable-Hamiltonian-structure-by-Beltrami}
is $-n$ in this example. 
Hence the pair $(\star_{3}\bm{v},\bm{v})$ 
is a stable Hamiltonian structure on $T^{3}$, where
$$
\star_{3}\bm{v}_{n}
=c(\,\cos(nx^{3})\dr x^{2}\wedge\dr x^{3}+\sin(nx^{3})\dr x^{3}\wedge\dr x^{1}).
$$
The Reeb vector field $Y_{n}$ is obtained as 
$$
Y_{n}
=\frac{\cos(nx^{3})}{c}\frac{\partial}{\partial x^{1}}
+\frac{\sin(nx^{3})}{c}\frac{\partial}{\partial x^{2}}.
$$
In addition, for the stable Hamiltonian structure
$(\star_{3}\bm{v},\bm{v}/\bm{g}^{-1}(\bm{v},\bm{v}))$, the Reeb vector field
is
$$
Z_{n}
=\wt{\bm{v}}^{\bm{g}}
=c\cos(nx^{3})\frac{\partial}{\partial x^{1}}
+c\sin(nx^{3})\frac{\partial}{\partial x^{2}}.
$$
\end{Example}
\begin{Example}
Consider the $3$-torus $\cN=T^{3}$, and 
the metric $\bm{g}$ is induced from the standard one on $\mbbR^{3}$. 
Let $\bm{v}$ be the $1$-form such that
$$
\bm{v}
=(A\sin x^{3}+C\cos x^{2})\dr x^{1}
+(B\sin x^{1}+A\cos x^{3})\dr x^{2}
+(C\sin x^{2}+B\cos x^{1})\dr x^{3},  
$$
with $A,B$, and $C$ being constant. The metric dual of $\bm{v}$, 
$Z=\wt{\bm{v}}^{\bm{g}}$, is called 
the {\it Arnold–Beltrami–Childress (ABC) flow} 
(see \cite{Dombre1986} for applications in fluid mechanics).  
Straightforward calculations yield 
\beqa
\bm{\dr}\bm{v}
&=&(A\sin x^{3}+C\cos x^{2})\dr x^{2}\wedge\dr x^{3}
+(B\sin x^{1}+A\cos x^{3})\dr x^{3}\wedge\dr x^{1}
+(C\sin x^{2}+B\cos x^{1})\dr x^{1}\wedge\dr x^{2},
\non\\
\star_{3}\bm{\dr}\bm{v}
&=&(A\sin x^{3}+C\cos x^{2})\dr x^{1}
+(B\sin x^{1}+A\cos x^{3})\dr x^{2}
+(C\sin x^{2}+B\cos x^{1})\dr x^{3},
\non
\eeqa
from which, it follows that $\star_{3}\bm{\dr}\bm{v}=\bm{v}$.
This shows that 
$k$ in  Proposition\,\ref{fact-stable-Hamiltonian-structure-by-Beltrami}
is $1$ in this example. Hence the pair $(\star_{3}\bm{v},\bm{v})$ 
is a stable Hamiltonian structure on $T^{3}$, where
$$
\star_{3}\bm{v}
=(A\sin x^{3}+C\cos x^{2})\dr x^{2}\wedge\dr x^{3}
+(B\sin x^{1}+A\cos x^{3})\dr x^{3}\wedge\dr x^{1}
+(C\sin x^{2}+B\cos x^{1})\dr x^{1}\wedge\dr x^{2}.
$$
Then, the Reeb vector field $Y$ is given by
$$
Y=\frac{A\sin x^{3}+C\cos x^{2}}{\bm{g}^{-1}(\bm{v},\bm{v})}
\frac{\partial}{\partial x^{1}}
+\frac{B\sin x^{1}+A\cos x^{3}}{\bm{g}^{-1}(\bm{v},\bm{v})}
\frac{\partial}{\partial x^{2}}
+\frac{C\sin x^{2}+B\cos x^{1}}{\bm{g}^{-1}(\bm{v},\bm{v})}
\frac{\partial}{\partial x^{3}},
$$
with 
$$
\bm{g}^{-1}(\bm{v},\bm{v})
=(A\sin x^{3}+C\cos x^{2})^{2}
+(B\sin x^{1}+A\cos x^{3})^{2}
+(C\sin x^{2}+B\cos x^{1})^{2}.
$$
In addition, for the stable Hamiltonian structure
$(\star_{3}\bm{v},\bm{v}/\bm{g}^{-1}(\bm{v},\bm{v}))$, the Reeb vector field
is
$$
Z=\wt{\bm{v}}^{\bm{g}}
=(A\sin x^{3}+C\cos x^{2})\frac{\partial}{\partial x^{1}}
+(B\sin x^{1}+A\cos x^{3})\frac{\partial}{\partial x^{2}}
+(C\sin x^{2}+B\cos x^{1})\frac{\partial}{\partial x^{3}}.
$$
\end{Example}
\begin{Example}
  Consider the solid torus $D^{2}\times S^{1}$. 
The coordinates for $D^{2}$ are set to be $r$ and $\varphi$ so that 
$(x^{1},x^{2})=(r\cos\varphi,r\sin\varphi)$, where  
$r^{2}=(x^{1})^{2}+(x^{2})^{2}$, 
  $D^{2}=\{(r,\varphi)\,|\,0<r\leq a,\ 0\leq\varphi<2\pi  \}$ with fixed $a>0$, 
and the coordinate for $S^{1}$ is set to be $x^{3}$. 
The metric tensor field is chosen as 
$\bm{g}$ that is the induced metric from
the standard metric of $\mbbR^{3}$:
$$
\bm{g}
=\dr r\otimes\dr r+r^{2}\dr \varphi\otimes \dr\varphi
+\dr x^{3}\otimes\dr x^{3}.
$$
The actions of the Hodge map are as follows:
$$
\star_{3}\dr r
=r\dr\varphi\wedge\dr x^{3},\quad
\star_{3}(r\dr\varphi)
=\dr x^{3}\wedge\dr r,\quad
\star_{3}\dr x^{3}
=\dr r\wedge r\dr\varphi,
$$
and
$$
\star_{3}(r\dr \varphi\wedge\dr x^{3})
=\dr r,\quad
\star_{3}(\dr x^{3}\wedge\dr r)
=r\dr \varphi,\quad
\star_{3}(\dr r\wedge r\dr\varphi)
=\dr x^{3}.
$$
Let $\bm{v}_{-}$ and $\bm{v}_{+}$ be the $1$-forms:
\beqa
\bm{v}_{-}
&=&\frac{\beta}{k_{c}}J_{1}(k_{c}r)\sin(\beta x^{3})\dr r
-\frac{k}{k_{c}}J_{1}(k_{c}r) \cos(\beta x^{3})r \dr \varphi
+J_{0}(k_{c}r)\cos(\beta x^{3})\dr x^{3},
\non\\
\bm{v}_{+}
&=&\frac{\beta}{k_{c}}J_{1}(k_{c}r)\sin(\beta x^{3})\dr r
+\frac{k}{k_{c}}J_{1}(k_{c}r) \cos(\beta x^{3})r \dr \varphi
+J_{0}(k_{c}r)\cos(\beta x^{3})\dr x^{3},
\non
\eeqa
where $J_{n}(z)$ is the $n$-th Bessel function of the first kind, 
 $k_{c}>0$ and $\beta$ are constant, 
and put $k^{2}=\beta^{2}+k_{c}^{2}$.  
Since
$$
\bm{g}^{-1}
=\frac{\partial}{\partial r}\otimes\frac{\partial}{\partial r}
+\frac{1}{r^{2}}
\frac{\partial}{\partial\varphi}\otimes\frac{\partial}{\partial\varphi}
+\frac{\partial}{\partial x^{3}}\otimes\frac{\partial}{\partial x^{3}},
$$
one has that
$$
\bm{g}^{-1}(\bm{v}_{-},\bm{v}_{-})
=\bm{g}^{-1}(\bm{v}_{+},\bm{v}_{+})
=\beta^{2}\left(\frac{J_{1}(k_{c}r)}{k_{c}}\right)^{2}
+\left(J_{0}(k_{c}r)+J_{1}(k_{c}r)\right)\cos^{2}(\beta x^{3}).
$$
There are several formulae for the Bessel functions, including
$$
\frac{\dr }{\dr z}J_{0}(z)
=-J_{1}(z),\qquad\text{and}\qquad
\frac{\dr }{\dr z}J_{1}(z)
=J_{0}(z)-\frac{1}{z}J_{1}(z),
$$
which yield 
$$
\frac{\dr }{\dr r}\left(rJ_{1}(k_{c}r)\right)
=rk_{c}J_{0}\left(k_{c}r\right).
$$
From these expressions, one has that
$$
k_{c}\bm{\dr} \bm{v}_{\mp}
=(\beta^{2}+k_{c}^{2})J_{1}(k_{c}r)
\cos(\beta x^{3})\dr x^{3}\wedge \dr r
\pm k [\beta J_{1}(k_{c}r)\sin(\beta x^{3})\dr x^{3}
-k_{c}J_{0}(k_{c}r)\cos(\beta x^{3})\dr r
]\wedge r\dr \varphi,
$$
yielding 
$$
k_{c}\star_{3}\bm{\dr} \bm{v}_{\mp}
=k^{2}J_{1}(k_{c}r)
\cos(\beta x^{3})r\dr \varphi
\pm k [-\beta J_{1}(k_{c}r)\sin(\beta x^{3})\dr r 
-k_{c}J_{0}(k_{c}r)\cos(\beta x^{3})\dr x^{3}
].
$$
Hence, one has that 
$$
\star_{3}\bm{\dr} \bm{v}_{-}
=- k \bm{v}_{-},\qquad\text{and}\qquad
\star_{3}\bm{\dr} \bm{v}_{+}
=+ k \bm{v}_{+}.
$$
Therefore the pairs
$(\star_{3}\bm{v}_{-},\bm{v}_{-})$
and $(\star_{3}\bm{v}_{+},\bm{v}_{+})$ are 
stable Hamiltonian structures. The Reeb vector fields
$Y_{-}$ and $Y_{+}$ are
$Y_{-}=Z_{-}/\bm{g}^{-1}(\bm{v}_{-},\bm{v}_{-})$ and
$Y_{+}=Z_{+}/\bm{g}^{-1}(\bm{v}_{+},\bm{v}_{+})$, where 
\beqa
Z_{-}
&=&\frac{\beta}{k_{c}}J_{1}(k_{c}r)\sin(\beta x^{3})\frac{\partial}{\partial r}
-\frac{k}{k_{c}}J_{1}(k_{c}r) \cos(\beta x^{3})\frac{1}{r}
\frac{\partial}{\partial \varphi}
+J_{0}(k_{c}r)\cos(\beta x^{3})\frac{\partial}{\partial x^{3}},
\non\\
Z_{+}
&=&\frac{\beta}{k_{c}}J_{1}(k_{c}r)\sin(\beta x^{3})\frac{\partial}{\partial r}
+\frac{k}{k_{c}}J_{1}(k_{c}r) \cos(\beta x^{3})\frac{1}{r}
\frac{\partial}{\partial \varphi}
+J_{0}(k_{c}r)\cos(\beta x^{3})\frac{\partial}{\partial x^{3}}.
\non
\eeqa
See \cite{Mochizuki2022} for applications of these fields in electromagnetism. 
\end{Example}

\section{Geometric properties of electromagnetic Beltrami fields }
\label{section-electromagnetic-Beltrami}
In this section, geometric properties of Maxwell fields are
discussed by means of notions discussed 
in Section\,\ref{section-preliminaries}.
To show such properties, 
let $\cN_{x^{0}}$ denote the hypersurface in $\cM$ at a fixed $x^{0}$.    
We then focus on the following 
at a fixed time.
\begin{Definition}
\label{definition-field-line}
Let $x^{0}$ be fixed.
Consider Maxwell fields on a $3$-dimensional
Riemannian manifold $(\cN_{x^{0}},\bm{g})$.
{\it Field lines of $\bm{e}$} at $x^{0}$ are integral curves of
the vector field $\wt{\bm{e}}^{\bm{g}}$
that is obtained by the metric dual of $\bm{e}$.
Similarly {\it field lines of $\bm{h}$} at
$x^{0}$ are integral curves of
the vector field $\wt{\bm{h}}^{\bm{g}}$.
\end{Definition}
Field lines in Definition\,\ref{definition-field-line}
are studied in this section,
and these integral curves are identified with orbits at $x^{0}$. 
Typical closed 
field lines are depicted in Figure\,\ref{picture-field-lines-e-h} 
in Section\,\ref{section-introduction}.  
A simple example of non-closed field lines is given as follows. 
\begin{Example}
\label{example-3-torus-SHS}
Let $(\cN_{x^{0}},\bm{g})$ be the Riemannian manifold with
$\cN_{x^{0}}=\mbbR^{3}$ and $\bm{g}_{\mbbR^{3}}$ given by \fr{g-E^3}.  
In addition, let $e_{0}$ and $h_{0}$ be positive constant, and 
$$
\bm{B}
=\mu_{0}h_{0}\dr x^{2}\wedge \dr x^{3},\quad
\bm{D}
=\varepsilon_{0}e_{0}\dr x^{2}\wedge \dr x^{3},\quad
\bm{e}
=e_{0}\dr x^{1},\quad
\bm{h}
=h_{0}\dr x^{1}.
$$
By substituting these form fields into \fr{Maxwell-decomposed-vacuo},
one verifies that they are Maxwell fields.
Field lines of $\bm{e}$ at $x^{0}$ are obtained as follows.
First, the metric dual of $\bm{e}$ 
is $\wt{\bm{e}}^{\bm{g}}=e_{0}\partial/\partial x^{1}$. Second, 
the integral curves of the field lines are obtained by solving 
the system of ODEs: 
$$
\frac{\dr x^{1}}{\dr s}
=e_{0},\qquad
\frac{\dr x^{2}}{\dr s}
=0,\qquad
\frac{\dr x^{3}}{\dr s}
=0,\qquad s\in\mbbR.
$$
Finally, the field lines are expressed as the solutions given by  
$$
x^{1}(s)
=e_{0}s+x^{1}(0),\qquad
x^{2}(s)
=x^{2}(0),\qquad
x^{3}(s)
=x^{3}(0).
$$
Note that these integral curves are not closed, and $s$ is not time, but $s$ 
represents a parameter of the curves. 
\end{Example}

Maxwell's equations in this section are assumed to
describe electromagnetic fields in vacuo unless otherwise stated. 

\subsection{Symplectic and stable Hamiltonian structures for Maxwell fields}
\label{section-symplectic-SHS}
Our strategy to draw connections between Maxwell fields 
and contact topology is to see if there is a closed field
line of $\bm{e}$ and that of $\bm{h}$, 
as mentioned in Section\,\ref{section-introduction}.  
To this end, one particular important class of Maxwell fields is chosen, and  
relations between 
field lines and Reeb vector fields are shown later.  

Before focusing specific Maxwell fields, basic
geometric properties of  
general Maxwell fields in vacuo are stated. 
\begin{Lemma}
\label{fact-F_0-F_1-symplectic}
Let $\bm{B},\bm{D},\bm{e},\bm{h}$ be solutions to Maxwell's equations on $\cM$. 
A Maxwell $2$-form $F_{0}$ given by \fr{F0} is symplectic if 
$$
\bm{B}\wedge\bm{e}\neq 0\qquad\text{on $\cM$}, 
$$
and $F_{1}$ given by \fr{F1} is symplectic if
$$
\bm{D}\wedge\bm{h}\neq 0\qquad\text{on $\cM$}. 
$$
\end{Lemma}
\begin{Proof}
Substituting \fr{F0} into $F_{0}\wedge F_{0}$ 
and recalling from  
Section\,\ref{section-preliminaries-Maxwell-equation} that 
$\ii_{\wt{\dr x^{0}}}\bm{B}=\ii_{\wt{\dr x^{0}}}\bm{e}=0$, one has that 
$$
F_{0}\wedge F_{0}
=c_{0}^{2}\bm{B}\wedge\bm{B}+2c_{0}\bm{B}\wedge\bm{e}\wedge\dr x^{0}.
$$
Since $\bm{B}$ is written of the form 
$\bm{B}=B_{3}\dr x^{1}\wedge\dr x^{2}+B_{1}\dr x^{2}\wedge\dr x^{3} 
+B_{2}\dr x^{3}\wedge\dr x^{1}$ with $B_{a}$, $(a=1,2,3)$ 
being some functions on $\cM$, one has $\bm{B}\wedge\bm{B}=0$. 
From this, if the given condition for $F_{0}$ is satisfied, then
$F_{0}\wedge F_{0}\neq 0$. 
In addition, since $\bm{B}$ and $\bm{e}$ are solutions to 
Maxwell's equations,  one has that $F_{0}$ is closed, $\dr F_{0}=0$.  
These show that $F_{0}$ is symplectic 
if the condition holds.  
A way to prove for $F_{1}$ is analogous to the proof for $F_{0}$. 
\qed
\end{Proof}

The $2$-forms $F_{0}$ and $F_{1}$ in Example\,\ref{Example-traveling-solution}
are not symplectic. Meanwhile the following $F_{0}$ and $F_{1}$ are symplectic.
\begin{Example}
As shown in Example\,\ref{example-3-torus-SHS},  
the $1$-form and $2$-form fields given by
$$
\bm{e}=e_{0}\dr x^{1},\quad
\bm{h}=h_{0}\dr x^{1},\quad
\bm{D}=\varepsilon_{0}\star_{3}\bm{e},\quad
\bm{B}=\mu_{0}\star_{3}\bm{h},  
$$
are Maxwell, where $e_{0}$ and $h_{0}$ are positive constant.  
In addition $F_{0}$ and $F_{1}$ are symplectic, due to
$$
\bm{B}\wedge\bm{e}
=\mu_{0}e_{0}h_{0}
\dr x^{1}\wedge\dr x^{2}\wedge\dr x^{3}\neq 0,\quad
\text{and}\quad
\bm{D}\wedge\bm{h}
=\varepsilon_{0}e_{0}h_{0}
\dr x^{1}\wedge\dr x^{2}\wedge\dr x^{3}
\neq 0.
$$
\end{Example}
Recall from Section\,\ref{section-preliminaries-Maxwell-equation} that   
$F_{0}$ and $F_{1}$ are defined on a Lorentzian manifold $(\cM,g)$, where  
the metric $g$ is of the form $g=-\dr x^{0}\otimes\dr x^{0}+\bm{g}$. 
Recall also that $\cN_{x^{0}}$ denotes the hypersurface in $\cM$ 
at a fixed $x^{0}$.  
The conditions that $F_{0}$ and $F_{1}$ are symplectic on $\cM$ are
strong, as has been shown in Example\,\ref{Example-traveling-solution}. 
Hence it is natural to consider weaker conditions for $F_{0}$ and $F_{1}$, so
that various classes of Maxwell fields can be considered. Such relaxed
structures are based on stable Hamiltonian structures  
(see Definition\,\ref{definition-stable-Hamiltonian-structure}). 
The following is associated with
Definition\,\ref{definition-stable-Hamiltonian-structure}.
\begin{Definition}
\label{definition-stable-Hamiltonian-structure-Maxwell}  
Let $F_{0}$ and $F_{1}$ be Maxwell fields on $\cI\times\cN$.  
Fix $x^{0}$, and let $\bm{g}$ be a Riemannian metric on 
$\cN_{x^{0}}$. 
If $\bm{B}$ is closed in the sense of $\bm{\dr}\bm{B}=0$ and 
$$
\bm{B}\wedge\bm{e}\neq 0,
\qquad\text{and}\qquad
\ker\bm{B}\subset\ker\bm{\dr}\bm{e},
$$
then the pair $(\bm{B},\bm{e})$ is called 
a {\it stable Hamiltonian structure for $F_{0}$  
on $\cN_{x^{0}}$}. Similarly,    
if $\bm{D}$ is closed in the sense of $\bm{\dr}\bm{D}=0$ 
and 
$$
\bm{D}\wedge\bm{h}\neq 0,
\qquad\text{and}\qquad
\ker\bm{D}\subset\ker\bm{\dr}\bm{h},
$$
then the pair $(\bm{D},\bm{h})$ is called
a {\it stable Hamiltonian structure for $F_{1}$ 
on $\cN_{x^{0}}$}. 
\end{Definition}

Given stable Hamiltonian structures for $F_{0}$ and $F_{1}$, 
there are corresponding Reeb vector fields $Y_{0}$ and $Y_{1}$
on $\cN$, defined such that
$$
\ii_{Y_{0}}\bm{B}
=0,\qquad
\ii_{Y_{0}}\bm{e}
=1,
$$
and
$$
\ii_{Y_{1}}\bm{D}
=0,\qquad
\ii_{Y_{1}}\bm{h}
=1.
$$
These Reeb vector fields preserve Maxwell fields in the following sense. 
\begin{Proposition}
\label{fact-conserved-energies-along-Reeb-SHS-general}  
Fix $x^{0}$. Then 
let $(\bm{B},\bm{e})$ be a stable Hamiltonian structure for $F_{0}$ 
on $\cN_{x^{0}}$.
The Reeb vector field $Y_{0}$ on $\cN$ 
preserves $\bm{e}$ and $\bm{B}$ in the sense that
$$
\cL_{Y_{0}}\bm{e}
=0,\qquad\text{and}\qquad
\cL_{Y_{0}}\bm{B}
=0.
$$
Similarly, let $(\bm{D},\bm{h})$ be a stable Hamiltonian structure for $F_{1}$
on $\cN_{x^{0}}$. 
The Reeb vector field $Y_{1}$ on $\cN_{x^{0}}$ 
preserves $\bm{h}$ and $\bm{D}$ in the sense that
$$
\cL_{Y_{1}}\bm{h}
=0,\qquad\text{and}\qquad
\cL_{Y_{1}}\bm{D}
=0.
$$
In addition, if there is a non-vanishing function $f$ such that
$Y_{0}=fY_{1}$, then it follows that 
$$
\cL_{Y_{0}}\bm{\cE}_{e}^{\kappa}
=0,\qquad
\cL_{Y_{1}}\bm{\cE}_{h}^{\kappa}
=0,
$$
where $\bm{\cE}_{e}^{\kappa}$ and $\bm{\cE}_{h}^{\kappa}$ 
have been defined in \fr{energy-densities-general-1}.
\end{Proposition}
A proof of
Proposition\,\ref{fact-conserved-energies-along-Reeb-SHS-general}
is given below and is similar to the proof for Item 2 of 
Remark\,\ref{remark-stable-Hamiltonian-structure}. The present proof
is written in terms of the Maxwell field. 
\begin{Proof}
There exists a non-vanishing function $f^{B}$ such that
$\bm{\dr}\bm{e}=f^{B}\bm{B}$ due to $\ker\bm{B}\subset\ker\bm{\dr}\bm{e}$.
With this and the conditions $\ii_{Y_{0}}\bm{e}=1$ and $\ii_{Y_{0}}\bm{B}=0$,
one has that 
$$
\cL_{Y_{0}}\bm{e}
=(\ii_{Y_{0}}\bm{\dr}+\bm{\dr}\ii_{Y_{0}})\bm{e}
=\ii_{Y_{0}}\bm{\dr}\bm{e}
=f^{B}\ii_{Y_{0}}\bm{B}
=0. 
$$
With $\ii_{Y_{0}}\bm{B}=0$ and the closedness condition 
(or the part of Maxwell's equations)  
$\bm{\dr}\bm{B}=0$, one has that 
$$
\cL_{Y_{0}}\bm{B}
=(\ii_{Y_{0}}\bm{\dr}+\bm{\dr}\ii_{Y_{0}})\bm{B}
=0.
$$
Similarly it follows that $\cL_{Y_{1}}\bm{h}=0$ and $\cL_{Y_{1}}\bm{D}=0$.
In addition, 
if there is a non-vanishing function $f$ such that $Y_{0}=fY_{1}$,   
it follows from 
$$
\cL_{Y_{0}}\bm{e}=0,\qquad\text{and}\qquad
\cL_{Y_{0}}\bm{D}
=(\bm{\dr}\ii_{Y_{0}}+\ii_{Y_{0}}\bm{\dr})\bm{D}
=\bm{\dr}\ii_{Y_{0}}\bm{D}
=\bm{\dr}(f\ii_{Y_{1}}\bm{D})
=0,
$$
that 
$$
\cL_{Y_{0}}\bm{\cE}_{e}^{\kappa}
=\frac{1}{2}(\cL_{Y_{0}}\bm{e})\wedge\bm{D}
+\frac{1}{2}\bm{e}\wedge(\cL_{Y_{0}}\bm{D})
=0.
$$
Similarly, one has that 
$$
\cL_{Y_{1}}\bm{\cE}_{h}^{\kappa}
=\frac{1}{2}(\cL_{Y_{1}}\bm{h})\wedge\bm{B}+
\frac{1}{2}\bm{h}\wedge(\cL_{Y_{1}}\bm{B})
=0.
$$
\qed
\end{Proof}

\begin{Example}
Consider the Maxwell fields given in Example\,\ref{example-3-torus-SHS}. 
Then, at a fixed $x^{0}$, 
the pair $(\bm{B},\bm{e})$ is a stable Hamiltonian structure
for $F_{0}$ on $\cN_{x^{0}}$,   
and the pair $(\bm{D},\bm{h})$ is that for $F_{1}$. 
In addition, the corresponding Reeb vector fields $Y_{0}$ and $Y_{1}$
on $\cN$ are expressed as
$$
Y_{0}
=\frac{1}{e_{0}}\frac{\partial}{\partial x^{1}},\qquad\text{and}\qquad
Y_{1}
=\frac{1}{h_{0}}\frac{\partial}{\partial x^{1}}, 
$$
respectively. Moreover the energy density forms for vacuo  
\fr{energy-densities} are expressed as    
$$
\bm{\cE}_{e}
=\frac{\varepsilon_{0}e_{0}^{2}}{2}\dr x^{1}\wedge\dr x^{2}\wedge \dr x^{3},
\qquad
\bm{\cE}_{h}
=\frac{\mu_{0}h_{0}^{2}}{2}\dr x^{1}\wedge\dr x^{2}\wedge \dr x^{3}.
$$
Then it can be verified for this case that 
$\cL_{Y_{0}}  \bm{\cE}_{e}=0$ and
$\cL_{Y_{1}}  \bm{\cE}_{h}=0$.  
\end{Example}

In Definition\,\ref{definition-stable-Hamiltonian-structure-Maxwell},
the pairs $(\bm{B},\bm{e})$ for $F_{0}$
and $(\bm{D},\bm{h})$ for $F_{1}$ on $\cN_{x^{0}}$ are considered. 
The closedness conditions for $\bm{B}$ and $\bm{D}$ are half of
Maxwell's equations in the decomposed form \fr{Maxwell-decomposed-vacuo}.
The other half of Maxwell equations involve time-derivatives of fields, and
to take into account these equations, $x^{0}$ should not be fixed.  
To deal with this, one considers
$(x_{0}^{0}-\epsilon,x_{0}^{0}+\epsilon)\subset\mbbR$ with some $\epsilon>0$
and a fixed $x_{0}^{0}$ 
such that fields can be differentiated with respect to $x^{0}$. 
This induces $\cM_{\epsilon}:=(x_{0}^{0}-\epsilon,x_{0}^{0}+\epsilon)\times\cN$ 
and then one has the symplectic manifolds
$(\cM_{\epsilon},F_{0})$ and
$(\cM_{\epsilon},F_{1})$ under some conditions, where
such symplectic manifolds have been shown in
Lemma\,\ref{fact-F_0-F_1-symplectic}. 
In addition, 
the conditions for   manifolds with stable Hamiltonian structures 
in Definition\,\ref{definition-stable-Hamiltonian-structure-Maxwell}
regarding
$\ker\bm{B}$ and $\ker\bm{D}$ restrict solutions to Maxwell's equations. 
Then it will be shown in
Proposition\,\ref{fact-stable-Hamiltonian-structure-Maxwell-fixed-time}  
that 
stable Hamiltonian structures $(\bm{B},\bm{e})$ and $(\bm{D},\bm{h})$
are appropriate for discussing a class of time-dependent Maxwell's fields. 
Furthermore, there are some cases where $F_{0}$ and $F_{1}$ are symplectic.   
As will be shown in
Theorem\,\ref{fact-twisted-mode-geometry-summery},
where a particular pair of $F_{0}$ and $F_{1}$ will be chosen,  
these time-dependent Maxwell fields do not yield symplectic forms
$F_{0}$ and $F_{1}$ on $\mbbR\times\cN$, however they do on $\cM_{\epsilon}$. 
 
In this subsection so far
symplectic manifolds and stable Hamiltonian structures
for Maxwell fields have been discussed. 
Before closing this subsection the 
applicability of the standard contact geometric methods to Maxwell's equations 
is briefly discussed. 
As shown below, given time-independent Maxwell  
fields, $\bm{e}$ and $\bm{h}$ cannot be contact forms. 
\begin{Proposition}
\label{fact-time-independent-Maxwell-no-contact}
Let $\cM_{\cI_{\epsilon}}:=\bigcup_{x^{0}\in \cI_{\epsilon}}
(\{x^{0}\}\times\cN_{x^{0}})$, and  
$\{(\cN_{x^{0}},\bm{g})\}_{x^{0}\in \cI_{\epsilon}}$
be a Riemannian manifold family.
Then time-independent Maxwell fields in vacuo
on $\cM_{\cI_{\epsilon}}$ do not induce 
contact manifolds with the contact forms $\bm{e}$ and $\bm{h}$
on $\cN_{x^{0}}$ at each fixed $x^{0}$.
\end{Proposition}
\begin{Proof}
Since the fields are time-independent and satisfy Maxwell's equations
\fr{Maxwell-decomposed-vacuo}, one has that 
$$
\bm{\dr}\bm{e}
=-\cL_{\partial/\partial t}\bm{B}
=0,\qquad\text{ and}\qquad 
\bm{\dr}\bm{h}
=\cL_{\partial/\partial t}\bm{D}
=0,\qquad\text{on $\cM_{\cI_{\epsilon}}$}.
$$
Hence
$$
\bm{e}\wedge\bm{\dr}\bm{e}
=0,\qquad\text{and}\qquad
\bm{h}\wedge\bm{\dr}\bm{h}
=0,  \qquad\text{on $\cN_{x^{0}}$ for each fixed $x^{0}$}. 
$$
These mean that $\bm{e}$ and $\bm{h}$ are not contact forms on $\cN_{x^{0}}$. 
\qed
\end{Proof}
Note that the trivial bundle $\cM_{\cI_{\epsilon}}$  
in Proposition\,\ref{fact-time-independent-Maxwell-no-contact} is 
equivalent to $\cI_{\epsilon}\times\cN$. 
The use of the notation $\cM_{\cI_{\epsilon}}$ is to emphasize 
that at each $x^{0}$, $\cN_{x^{0}}$ is a contact manifold. 

\subsection{Contact forms and stable Hamiltonian structures for $F_{0}$ and $F_{1}$}
  In this subsection stable Hamiltonian structures for $F_{0}$ and $F_{1}$
  on $(\cM,g)$ in Section\,\ref{section-symplectic-SHS} 
  are explicitly constructed. This construction
  is achieved by introducing a $1$-form $\wh{\bm{v}}$ on $\cM$, where 
  $\wh{\bm{v}}$ is analogous to a    
  rotational Beltrami $1$-form $\bm{v}$ on $(\cN,\bm{g})$. This $\bm{v}$ has
  been introduced in Section\,\ref{section-geometry-Beltrami},  
  and the relation between involved manifolds is
  $\cM=\cI\times\cN$
  and the relation between
  involved metrics is $g=-\dr x^{0}\otimes\dr x^{0}+\bm{g}$.   
  After constructing Maxwell fields with $\wh{\bm{v}}$, 
  the pairs  
  $(\bm{B},\bm{e})$ and $(\bm{B},\bm{h})$
  are shown to be stable Hamiltonian structures. 
  In addition, $\bm{e}$ and $\bm{h}$ are shown to be contact forms.
  Their properties are then discussed. 

  The point of departure in this subsection is to
  introduce a Hodge map like operator $\wh{\star_{3}}$ on $\cM$,  because 
  the definition of $\wh{\bm{v}}$ involves this operator. 
  To define $\wh{\star_{3}}$ on $(\cM,g)$, write 
  $\bm{g}=\sum_{a=1}^{3}\theta^{a}\otimes\theta^{a}$,  where  
  $\cL_{\partial/\partial x^{0}}\theta^{a}=0$, $a=1,2,3$. 
  Let $\wh{\star_{3}}1$ be the $3$-form on $\cM$   
  $$
  \wh{\star_{3}}1
  =\theta^{1}\wedge\theta^{2}\wedge\theta^{3}. 
  $$
  The action of $\wh{\star_{3}}$ on $r$-forms on $\cM$ is defined such that
  $$
  \wh{\star_{3}}(\bm{\alpha}_{1}\wedge\cdots\wedge\bm{\alpha}_{r})
  =\ii_{\bm{g}^{-1}(\bm{\alpha}_{r},-)}\cdots\ii_{\bm{g}^{-1}(\bm{\alpha}_{1},-)}
  \wh{\star_{3}}1,\quad 1\leq r\leq 3,
  $$
  for all $\bm{\alpha}_{1},\bm{\alpha}_{2},\bm{\alpha}_{3}\in\GamLamM{1}$
  satisfying  
  $$
  \ii_{\partial/\partial x^{0}}\bm{\alpha}_{1}
  =\ii_{\partial/\partial x^{0}}\bm{\alpha}_{2}
  =\ii_{\partial/\partial x^{0}}\bm{\alpha}_{3}
  =0.
  $$ 
  This definition leads to the following: 
  $$
\wh{\star_{3}}\theta^{1}
  =\theta^{2}\wedge\theta^{3},\quad
  \wh{\star_{3}}\theta^{2}
  =\theta^{3}\wedge\theta^{1},\quad  
  \wh{\star_{3}}\theta^{3}
  =\theta^{1}\wedge\theta^{2},
$$
  $$
  \wh{\star_{3}}\left(\theta^{1}\wedge\theta^{2}\right)
  =\theta^{3},\quad
  \wh{\star_{3}}\left(\theta^{2}\wedge\theta^{3}\right)
  =\theta^{1},\quad  
  \wh{\star_{3}}\left(\theta^{3}\wedge\theta^{1}\right)
  =\theta^{2},
  $$
$$
  \wh{\star_{3}}\left(\theta^{1}\wedge\theta^{2}\wedge\theta^{3}\right)
  =1.
  $$
  From straightforward calculations, it follows that 
  $$
  \wh{\star_{3}}\wh{\star_{3}}\bm{\alpha}
  =\bm{\alpha},
  $$
  where $\bm{\alpha}$ is a $k$-form such
  that $\ii_{\partial/\partial x^{0}}\bm{\alpha}=0$, $k=0,1,2,3$.   
  Hence the action of $\wh{\star_{3}}$ is akin to that of $\star_{3}$, 
  where $\star_{3}$ has been defined on $\cN$ in 
  Section\,\ref{section-preliminaries}.  
  
  A $1$-form on $\cM$ denoted by $\wh{\bm{v}}$ plays a fundamental role 
  in this subsection, where $\wh{\bm{v}}$ 
  is an analogue of $\bm{v}$ in Section\,\ref{section-geometry-Beltrami}.
  A $1$-form $\wh{\bm{v}}$ is called a
  {\it rotational Beltrami $1$-form on $\cM$ with respect to $x^{0}$},  
  if all of the following are satisfied: 
  $$
  \wh{\star_{3}}\bm{\dr}\wh{\bm{v}}
  =\wh{f}\wh{\bm{v}},\quad
  \cL_{\partial/\partial x^{0}}\wh{\bm{v}}
  =0,\quad \text{and}\quad
  \ii_{\partial/\partial x^{0}}\wh{\bm{v}}
  =0,\qquad 
  \forall\, x^{0}\in\cI
  $$
  with some $\wh{f}\neq 0\in\GamLamM{0}$, where $\wh{\star_{3}}$
  has been defined above.
 This $\wh{\bm{v}}$ is also called a {\it rotational Beltrami $1$-form on $\cM$,
   and is called a {\it rotational Beltrami $1$-form} 
   by abuse of terminology of Definition\,\ref{definition-Beltrami-1-form-f}.} 
  Several properties of rotational Beltrami $1$-forms  
  on $\cM$ are shown below. 
  \begin{enumerate}
  \item
    It follows that 
    $$
    \ii_{\partial/\partial x^{0}}\bm{\dr}\wh{\bm{v}}
    =0,
    $$
    which is shown as
    $$
    0=\cL_{\partial/\partial x^{0}}\wh{\bm{v}}
    =\left(\dr\ii_{\partial/\partial x^{0}}+\ii_{\partial/\partial x^{0}}\dr\right)
    \wh{\bm{v}}
    =\ii_{\partial/\partial x^{0}}\dr\wh{\bm{v}}
    =\ii_{\partial/\partial x^{0}}\left(\bm{\dr}+\dr x^{0}\wedge\cL_{\partial/\partial x^{0}}\right)\wh{\bm{v}}
    =\ii_{\partial/\partial x^{0}}\bm{\dr}\wh{\bm{v}}.
    $$
\item
  It follows that 
  $$
  g^{-1}(\wh{\bm{v}},-)
  =\bm{g}^{-1}(\wh{\bm{v}},-)\quad \in\GTM,
  $$
  due to
  $\ii_{\partial/\partial x^{0}}\wh{\bm{v}}=0$.
  \item
    Consider a class of non-singular
    rotational Beltrami $1$-form $\wh{\bm{v}}$ on $\cM$  such that 
  $$
    \wh{\star_{3}}\bm{\dr}\wh{\bm{v}}
  =k\wh{\bm{v}},\quad
  \cL_{\partial/\partial x^{0}}\wh{\bm{v}}
  =0,\quad\text{and}\quad
  \ii_{\partial/\partial x^{0}}\wh{\bm{v}}
  =0, 
  \quad \forall\, x^{0}\in\cI,
  $$
  with $k\neq 0$ being a constant. By definition of non-singular $1$-form,
  $\wh{\bm{v}}$ is such that $\bm{g}^{-1}(\wh{\bm{v}},\wh{\bm{v}})\neq 0$. 
  Then $\wh{\bm{v}}$ satisfies
  $$
  \wh{\bm{v}}\wedge\wh{\star_{3}}\wh{\bm{v}}
  \neq 0,
  $$
  which shows that the pullback of $\wh{\bm{v}}$ 
  is a contact form on a submanifold $\cN_{x^{0}}$ at a fixed $x^{0}$.
  To prove that $\wh{\bm{v}}$ on $\cN_{x^{0}}$ is a contact form, the following two identities are employed:
\begin{enumerate}
\item
  The first identity is 
  $$
      g^{-1}(\wh{\bm{v}},\wh{\bm{v}})\wh{\star_{3}}1
  =\wh{\bm{v}}\wedge\ii_{g^{-1}(\wh{\bm{v}},-)}\wh{\star_{3}}1,
  $$
  which is obtained from 
  $$
  0=
\ii_{g^{-1}(\wh{\bm{v}},-)}\left(\wh{\bm{v}}\wedge\wh{\star_{3}}1\right)
=g^{-1}(\wh{\bm{v}},\wh{\bm{v}})\wh{\star_{3}}1
  -\wh{\bm{v}}\wedge\ii_{g^{-1}(\wh{\bm{v}},-)}\wh{\star_{3}}1.
  $$
\item
  The second identity is
  $$
\wh{\bm{v}}\wedge  \ii_{g^{-1}(\wh{\bm{v}},-)}\wh{\star_{3}}1
=\wh{\bm{v}}\wedge\wh{\star_{3}}\wh{\bm{v}}.
  $$
  The following is enough to consider: 
  $$
  \ii_{g^{-1}(\wh{\bm{v}},-)}\wh{\star_{3}}1
  =\wh{\star_{3}}\wh{\bm{v}}.
  $$
  To show this, write
  $$
  \wh{\bm{v}}
  =\sum_{a=1}^{3}\wh{v}_{a}\theta^{a}\quad \in\GamLamM{1}
  $$
  with $\wh{v}_{1},\wh{v}_{2},\wh{v}_{3}\in\GamLamM{0}$ satisfying  
  $\cL_{\partial/\partial x^{0}}\wh{v}_{a}=0$, $a=1,2,3$, so  that
  $\cL_{\partial/\partial x^{0}}\wh{\bm{v}}=0$. 
  Then one has  
  $$
  g^{-1}(\wh{\bm{v}},-)
  =\sum_{a=1}^{3}\wh{v}_{a}X_{a}  \quad\in\GTM.
  $$
  Using these expressions, one has that 
\beqa
  \ii_{g^{-1}(\wh{\bm{v}},-)}\wh{\star_{3}}1
  &=&\ii_{g^{-1}(\wh{\bm{v}},-)}\left(\theta^{1}\wedge\theta^{2}\wedge\theta^{3}\right)
  =\wh{v}_{1}\theta^{2}\wedge\theta^{3}
  +\wh{v}_{2}\theta^{3}\wedge\theta^{1}
  +\wh{v}_{3}\theta^{1}\wedge\theta^{2},
  \non\\
  \wt{\star_{3}}\wh{\bm{v}}
  &=&\wh{v}_{1}\theta^{2}\wedge\theta^{3}
  +\wh{v}_{2}\theta^{3}\wedge\theta^{1}
  +\wh{v}_{3}\theta^{1}\wedge\theta^{2}.
  \non
  \eeqa
  These two equations show that the second identity holds. 
\end{enumerate}
Combining the two identities above and 
$g(\wh{\bm{v}},\wh{\bm{v}})\neq 0$,
one has that 
$$
0\neq g(\wh{\bm{v}},\wh{\bm{v}})\wh{\star_{3}}1
=\wh{\bm{v}}\wedge \wh{\star_{3}}\wh{\bm{v}}
=\wh{\bm{v}}\wedge \left(k^{-1}\wh{\star_{3}}\bm{\dr}\wh{\bm{v}}\right).
$$
This leads to $\wh{\bm{v}}\wedge\wh{\star_{3}}\wh{\bm{v}}\neq0$. 
In addition, as can be shown, the pullback of $\wh{\bm{v}}$ onto $\cN_{x^{0}}$ 
has the properties stated in 
Proposition\,\ref{fact-stable-Hamiltonian-structure-by-Beltrami} and Corollary\,\ref{fact-stable-Hamiltonian-structure-by-Beltrami-additional}.   

\end{enumerate}
  A rotational Beltrami $1$-form $\bm{v}$ on $\cN_{x^{0}}$ at a fixed $x^{0}$ 
  is obtained by the pullback of $\wh{\bm{v}}$, 
  $\bm{v}=\iota_{x^{0}}^{*}\wh{\bm{v}}$, where $\iota_{x^{0}}:\cN_{x^{0}}\to\cM$
  denotes the embedding. 
  In what follows this $\wh{\bm{v}}$ is simply written as $\bm{v}$
  and the maps $\star_{3}$ and $\wh{\star_{3}}$
  are not distinguish by abuse of notation.
  In addition the pullback of forms is often omitted throughout.  

The following shows how to construct contact forms and 
stable Hamiltonian structures for $F_{0}$ and $F_{1}$.  

\begin{Proposition}
\label{fact-stable-Hamiltonian-structure-Maxwell-fixed-time}
Let $\bm{v}$ be a non-singular rotational Beltrami $1$-form
on $\cM$ such that 
$$
\star_{3}\bm{\dr}\bm{v}
=k\bm{v},
$$
with $k$ being a non-zero constant. In addition,   
let the forms 
\beq
\bm{e}=f^{e}\bm{v},\qquad
\bm{h}=f^{h}\bm{v},
\label{e-h-generated-by-the-same-v}
\eeq
\beq
\bm{B}=\mu_{0}\star_{3}\bm{h},\qquad
\bm{D}=\varepsilon_{0}\star_{3}\bm{e},
\label{B-D-generated-by-the-same-v}
\eeq 
be Maxwell fields, where 
$f^{e}$ and $f^{h}$ are non-vanishing functions of $x^{0}$ on some
domain 
so that $\bm{\dr} f^{e}=\bm{\dr}f^{h}=0$. 
Then the pairs $(\bm{B},\bm{e})$ and $(\bm{D},\bm{h})$
are stable Hamiltonian structures for $F_{0}$ and $F_{1}$ 
on $\cN_{x^{0}}$, respectively.
In addition, $\bm{e}$ and $\bm{h}$ are contact forms
on $\cN_{x^{0}}$ when $f^{e}$ and $f^{h}$ are non-vanishing, respectively.  
\end{Proposition}
\begin{Proof}
To prove the statement, the following are shown.
(i) $\bm{B}$ and $\bm{D}$ are closed, 
(ii) $\bm{B}\wedge\bm{e}\neq 0$ and 
$\bm{D}\wedge\bm{h}\neq 0$,  
(iii) 
$\ker\bm{B}\subset\ker\bm{\dr}\bm{e}$ and
$\ker\bm{D}\subset\ker\bm{\dr}\bm{h}$.
In addition,
(iv) $\bm{e}\wedge\bm{\dr}\bm{e}\neq 0$ and
$\bm{h}\wedge\bm{\dr}\bm{h}\neq 0$. 

(Proof for i). It follows from
$\bm{\dr}f^{e}=0$, $\bm{\dr}f^{h}=0$, 
\beq
\bm{B}
=\mu_{0}\star_{3}\bm{h}
=\mu_{0}\frac{f^{h}}{k}\bm{\dr}\bm{v},
\qquad\text{and}\qquad  
\bm{D}
=\varepsilon_{0}\star_{3}\bm{e}
=\varepsilon_{0}\frac{f^{e}}{k}\bm{\dr}\bm{v},
\label{B-D-stable-Hamiltonian-structure-Maxwell-fixed-time}
\eeq
that $\bm{B}$ and $\bm{D}$ are closed, $\bm{\dr}\bm{B}=\bm{\dr}\bm{D}=0$.

(Proof for ii). 
The non-singular condition for $\bm{v}$ translates into
$\bm{g}^{-1}(\bm{v},\bm{v})\neq 0$, from which one has that  
$$
\bm{v}\wedge\bm{\dr}\bm{v}
=k  \bm{v}\wedge\star_{3}\bm{v}  
=k\bm{g}^{-1}(\bm{v},\bm{v})\star_{3}1
\neq 0,\qquad\text{on $\cN_{x^{0}}$}.
$$
With this, \fr{e-h-generated-by-the-same-v}, and 
\fr{B-D-stable-Hamiltonian-structure-Maxwell-fixed-time}, one has that 
$$
\bm{B}\wedge\bm{e}
=\mu_{0}\frac{f^{e}f^{h}}{k}\bm{v}\wedge\bm{\dr}\bm{v}
\neq0,\quad\text{and}\quad
\bm{D}\wedge\bm{h}
=\varepsilon_{0}\frac{f^{e}f^{h}}{k}\bm{v}\wedge\bm{\dr}\bm{v}
\neq0,
\quad\text{on $\cN_{x^{0}}$}.
$$

(Proof for iii). 
The conditions 
$$
\ker\bm{B}\subset\ker\bm{\dr}\bm{e}
\qquad\text{and}\qquad
\ker\bm{D}\subset\ker\bm{\dr}\bm{h}
$$
are equivalent to 
$$
\bm{\dr}\bm{e}
=f^{B}\bm{B}
\qquad\text{and}\qquad
\bm{\dr}\bm{h}
=f^{D}\bm{D},  
$$
with $f^{B}$ and $f^{D}$ being some functions. 
The functions $f^{B}$ and $f^{D}$ are explicitly expressed as 
$$
f^{B}
=\frac{k}{\mu_{0}}\frac{f^{e}}{f^{h}},
\qquad\text{and}\qquad
f^{D}
=\frac{k}{\varepsilon_{0}}\frac{f^{h}}{f^{e}}.
$$
They are obtained as follows. From
\fr{B-D-stable-Hamiltonian-structure-Maxwell-fixed-time}    and 
\fr{e-h-generated-by-the-same-v} with 
$\bm{\dr} f^{e}=0$ and $\bm{\dr} f^{h}=0$, it follows that
$$
\bm{\dr}\bm{e}
=f^{e}\bm{\dr}\bm{v}
=\frac{k}{\mu_{0}}\frac{f^{e}}{f^{h}}\bm{B},
\qquad\text{and}\qquad
\bm{\dr}\bm{h}
=f^{h}\bm{\dr}\bm{v}
=\frac{k}{\varepsilon_{0}}\frac{f^{h}}{f^{e}}\bm{D}.  
$$

(Proof for iv).
As has been shown in Proof for ii, it follows that
$\bm{v}\wedge\bm{\dr}\bm{v}\neq 0$. With this,
\fr{e-h-generated-by-the-same-v}, and $f^{e}$ is non-zero,
one has that 
$$
\bm{e}\wedge\bm{\dr}\bm{e}
=(f^{e})^{2}\bm{v}\wedge\bm{\dr}\bm{v}
\neq 0.
$$
Similarly 
it follows that
$$
\bm{h}\wedge\bm{\dr}\bm{h}
\neq 0.
$$
Hence $\bm{e}$ and $\bm{h}$ are contact forms on $\cN_{x^{0}}$, 
when $f^{e}$ and $f^{h}$ are non-vanishing, respectively.    
\qed
\end{Proof}
The $x^{0}$ dependence for $f^{e}$ and $f^{h}$
in
Proposition\,\ref{fact-stable-Hamiltonian-structure-Maxwell-fixed-time} 
is determined by solving
the system of ODEs, 
$$
\frac{\dr f^{e}}{\dr x^{0}}
=\frac{k}{\varepsilon_{0}}f^{h},\qquad\text{and}\qquad
\frac{\dr f^{h}}{\dr x^{0}}
=-\frac{k}{\mu_{0}}f^{e},
$$
where this system has been obtained by substituting
\fr{e-h-generated-by-the-same-v} and \fr{B-D-generated-by-the-same-v}
into Maxwell's equations \fr{Maxwell-decomposed-vacuo}.  
A time-dependence of the Maxwell fields in
Proposition\,\ref{fact-stable-Hamiltonian-structure-Maxwell-fixed-time} 
will explicitly be fixed  
in Lemma\,\ref{fact-explicit-Maxwell-Beltrami-time-dependent}, where a
different convention of the sign for $k$ is adopted.

The Reeb vector fields
in Proposition\,\ref{fact-stable-Hamiltonian-structure-Maxwell-fixed-time} 
can be expressed in terms of  Maxwell fields.
\begin{Proposition}
\label{fact-stable-Hamiltonian-structure-Maxwell-Reeb}  
Let Maxwell fields be as in \fr{e-h-generated-by-the-same-v} and
\fr{B-D-generated-by-the-same-v},   
$(\bm{B},\bm{e})$ and $(\bm{D},\bm{h})$ be stable Hamiltonian structures 
for $F_{0}$ and $F_{1}$ on $\cN_{x^{0}}$, respectively. Then 
the Reeb vector fields $Y_{0}$ and $Y_{1}$ on $\cN_{x^{0}}$  
such that
$\ii_{Y_{0}}\bm{B}=0$, $\ii_{Y_{0}}\bm{e}=1$,
and
$
\ii_{Y_{1}}\bm{D}=0$, $\ii_{Y_{1}}\bm{h}=1$, 
are expressed as
\beq
Y_{0}
=\frac{\wt{\bm{e}}^{\bm{g}}}{\bm{g}^{-1}(\bm{e},\bm{e})}
\quad\text{and}\quad
Y_{1}
=\frac{\wt{\bm{h}}^{\bm{g}}}{\bm{g}^{-1}(\bm{h},\bm{h})}.
\label{stable-Hamiltonian-structure-Maxwell-Reeb-0-1-e-h}
\eeq
\end{Proposition}
\begin{Proof}
This statement can be verified by straightforward calculations.
Assume that $Y_{0}$ is given as above.
Substituting this $Y_{0}$ into $\ii_{Y_{0}}\bm{e}$,
one verifies that $\ii_{Y_{0}}\bm{e}=1$,  
$$
\ii_{Y_{0}}\bm{e}
= \bm{e}(Y_{0})
=\frac{\wt{\bm{e}}^{\bm{g}}(\bm{e})}{\bm{g}^{-1}(\bm{e},\bm{e})}
=1.
$$
Substituting this $Y_{0}$ into $\ii_{Y_{0}}\bm{B}$,
one verifies that $\ii_{Y_{0}}\bm{B}=0$,     
\beqa
\ii_{Y_{0}}\bm{B}
&=&\mu_{0}f^{h}\ii_{Y_{0}}\star_{3}\bm{v}
=\mu_{0}f^{h}\star_{3}\left(\bm{v}\wedge\wt{Y_{0}}^{\bm{g}}\right)
=\mu_{0}f^{h}\star_{3}
\left(\bm{v}\wedge\frac{f^{e}\bm{v}}{\bm{g}^{-1}(\bm{e},\bm{e})}\right)
\non\\
&=&\mu_{0}f^{h}f^{e}\frac{1}{\bm{g}^{-1}(\bm{e},\bm{e})}
\star_{3}(\bm{v}\wedge\bm{v})
=0.
\non
\eeqa
Hence, $Y_{0}$ satisfies the conditions of the Reeb vector field. 
Since the Reeb vector field is known to be unique, the proof for $Y_{0}$ is
completed. 
A way to prove for $Y_{1}$ is analogous to the proof for $Y_{0}$.
\qed
\end{Proof}

In Proposition\,\ref{fact-stable-Hamiltonian-structure-Maxwell-fixed-time},
there are two Reeb vector fields $Y_{0}$ and $Y_{1}$ due to the two 
stable Hamiltonian structures $(\bm{B},\bm{e})$ and $(\bm{D},\bm{h})$.
In general, these $Y_{0}$ and $Y_{1}$ need not be equal.  
To see a relation between $Y_{0}$ and $Y_{1}$, rewrite $Y_{0}$ and
$Y_{1}$ 
in Proposition\,\ref{fact-stable-Hamiltonian-structure-Maxwell-Reeb} as 
\beq
Y_{0}=\frac{\wt{\bm{v}}^{\bm{g}}}{f^{e}\bm{g}^{-1}(\bm{v},\bm{v})},\qquad
\text{and}\qquad
Y_{1}=\frac{\wt{\bm{v}}^{\bm{g}}}{f^{h}\bm{g}^{-1}(\bm{v},\bm{v})}.
\label{stable-Hamiltonian-structure-Maxwell-Reeb-0-1}
\eeq
These expressions immediately yield 
$$
Y_{1}
=\frac{f^{e}}{f^{h}}
Y_{0},
$$
from which $Y_{1}$ is (anti-)parallel to $Y_{0}$.  
In addition, at any point of $\cN$ if $f^{e}=f^{h}$, 
then it follows that $Y_{1}=Y_{0}$.
To study time-dependent Maxwell fields,
let $f^{e}$ and $f^{h}$ be some functions of $t$ or equivalently $x^{0}$
up to $c_{0}$,  
so that the conditions $\bm{\dr}f^{e}=\bm{\dr}f^{h}=0$ are satisfied. 
Then, for non-trivial time-dependent Maxwell fields on a $4$-dimensional
manifold $\cM_{\epsilon}$,  
the equality $f^{e}=f^{h}$ identically on $\cM_{\epsilon}$ is not realized,
as will be shown in
item 1 of Remark\,\ref{remark-Beltrami-parallel}.

The Reeb vector fields $Y_{0}$ and $Y_{1}$ have physically important
properties  
in the following sense.
\begin{Corollary}
\label{fact-conserved-energies-along-Reeb-SHS-Beltrami}  
Let $Y_{0}$ and $Y_{1}$
be as in Proposition\,\ref{fact-stable-Hamiltonian-structure-Maxwell-Reeb}.
Then the energy density forms $\bm{\cE}_{e}$ and $\bm{\cE}_{h}$
defined in \fr{energy-densities} are conserved along
$Y_{0}$ and $Y_{1}$, respectively, in the sense that
$$
\cL_{Y_{0}}\bm{\cE}_{e}
=\cL_{Y_{0}}\left(
\frac{\varepsilon_{0}}{2}\bm{e}\wedge\star_{3}\bm{e}\right)
=0,\qquad
\cL_{Y_{1}}\bm{\cE}_{e}
=\cL_{Y_{0}}\left(
\frac{\varepsilon_{0}}{2}\bm{e}\wedge\star_{3}\bm{e}\right)
=0,
$$
and
$$
\cL_{Y_{0}}\bm{\cE}_{h}
=\cL_{Y_{1}}\left(
\frac{\mu_{0}}{2}\bm{h}\wedge\star_{3}\bm{h}\right)
=0,
\qquad
\cL_{Y_{1}}\bm{\cE}_{h}
=\cL_{Y_{1}}\left(
\frac{\mu_{0}}{2}\bm{h}\wedge\star_{3}\bm{h}\right)
=0.  
$$
\end{Corollary}  
\begin{Proof}
The statement can be proved by straightforward calculations.
From $\ii_{Y_{0}}\bm{e}=1$, $\bm{e}=f^{e}\bm{v}$,
$\bm{\dr}\star_{3}\bm{e}=0$, and
$\bm{v}\wedge\wt{Y_{0}}^{\bm{g}}=0$ due to
\fr{stable-Hamiltonian-structure-Maxwell-Reeb-0-1}, it follows that 
\beqa
\cL_{Y_{0}}\bm{e}
&=&(\bm{\dr}\ii_{Y_{0}}+\ii_{Y_{0}}\bm{\dr})\bm{e}
=\bm{\dr} 1+f^{e}\ii_{Y_{0}}\bm{\dr}\bm{v}
=f^{e}k\ii_{Y_{0}}\star_{3}\bm{v}
=f^{e}k\star_{3}\left(\bm{v}\wedge \wt{Y_{0}}^{\bm{g}}\right)
=0,
\non\\
\cL_{Y_{0}}\star_{3}\bm{e}
&=&(\bm{\dr}\ii_{Y_{0}}+\ii_{Y_{0}}\bm{\dr})\star_{3}\bm{e}
=\bm{\dr}\ii_{Y_{0}}\star_{3}\bm{e}
=\bm{\dr}\star_{3}\left(\bm{e}\wedge\wt{Y_{0}}^{\bm{g}}\right)
=f^{e}\bm{\dr}\star_{3}\left(\bm{v}\wedge\wt{Y_{0}}^{\bm{g}}\right)
=0.
\non
\eeqa
With these equations, one has 
$$
\cL_{Y_{0}}\left(
\frac{\varepsilon_{0}}{2}\bm{e}\wedge\star_{3}\bm{e}\right)
=\frac{\varepsilon_{0}}{2}\left[
  (\cL_{Y_{0}}\bm{e})\wedge\star_{3}\bm{e}+\bm{e}\wedge
  (\cL_{Y_{0}}\star_{3}\bm{e})
  \right]
=0.
$$
From \fr{stable-Hamiltonian-structure-Maxwell-Reeb-0-1}, it follows that 
\beqa
\cL_{Y_{1}}\bm{e}
&=&f^{e}(\bm{\dr}\ii_{Y_{1}}+\ii_{Y_{1}}\bm{\dr})\bm{v}
=f^{e}(\bm{\dr}\ii_{Y_{1}}\bm{v}+\ii_{Y_{1}}\bm{\dr}\bm{v})
=f^{e}(\bm{\dr}\ii_{Y_{1}}\bm{v}+k\ii_{Y_{1}}\star_{3}\bm{v})
\non\\
&=&f^{e}\bm{\dr}\left(
\frac{\bm{g}^{-1}(\bm{v},\bm{v})}{f^{h}\bm{g}^{-1}(\bm{v},\bm{v})}\right)
+f^{e}k\star_{3}\left(\bm{v}\wedge\wt{Y_{1}}^{\bm{g}}\right)
=0,
\non\\
\cL_{Y_{1}}\star_{3}\bm{e}
&=&f^{e}(\bm{\dr}\ii_{Y_{1}}+\ii_{Y_{1}}\bm{\dr})\star_{3}\bm{v}  
=f^{e}\bm{\dr}\ii_{Y_{1}}\star_{3}\bm{v}
=f^{e}\bm{\dr}\star_{3}\left(\bm{v}\wedge\wt{Y_{1}}^{\bm{g}}\right)
=0.
\non
\eeqa
With these equations, one has 
$$
\cL_{Y_{1}}\left(\frac{\varepsilon_{0}}{2}\bm{e}\wedge\star_{3}\bm{e}\right)
=\frac{\varepsilon_{0}}{2}\left[
  (\cL_{Y_{1}}\bm{e})\wedge \star_{3}\bm{e}
  +\bm{e}\wedge(\cL_{Y_{1}}\star_{3}\bm{e})\right]
=0.
$$
A way to prove $\cL_{Y_{0}}\bm{\cE}_{h}=\cL_{Y_{1}}\bm{\cE}_{h}=0$ is analogous.   
\qed
\end{Proof}

As a consequence of 
Proposition\,\ref{fact-stable-Hamiltonian-structure-Maxwell-Reeb}, 
the following is observed. 
If $\bm{g}^{-1}(\bm{e},\bm{e})$ is constant, then 
the relation between $Y_{0}$ and $\bm{e}$ is one-to-one due to 
Proposition\,\ref{fact-stable-Hamiltonian-structure-Maxwell-Reeb}. 
In the case that the relation between $Y_{0}$ and $\bm{e}$ is one-to-one, 
if $Y_{0}$ generates a closed orbit, then a field line of $\bm{e}$ is 
closed. Similarly, in the case that the relation between 
$Y_{1}$ and $\bm{h}$ is one-to-one, if $Y_{1}$ generates a closed orbit, 
then a field line of $\bm{h}$ is closed.  
This argument can be refined, and will be stated in 
Theorem\,\ref{fact-Hutchings2009-electromagnetism}. 

Motivated by the shown several properties of the Maxwell fields  
in Proposition\,\ref{fact-stable-Hamiltonian-structure-Maxwell-fixed-time}, 
the following are defined in this paper.
\begin{Definition}
\label{definition-Maxwell-Beltrami} 
If $\bm{e}$ and $\bm{h}$ in Maxwell fields are proportional to a 
non-singular rotational Beltrami $1$-form $\bm{v}$, then 
Maxwell fields are called {\it Beltrami-Maxwell fields} in this paper.
\end{Definition}
\begin{Definition}
\label{definition-Maxwell-parallel}
If $\bm{e}$ and $\bm{h}$ in Maxwell fields satisfy  
$$
\bm{e}\wedge\bm{h}
=0\quad
\text{ on $\cM\setminus\partial\cM$,}  
$$
then Maxwell fields are called 
{\it $\bm{e}\paral\bm{h}$ fields}, or {\it parallel fields} in this paper.  
\end{Definition}
For all $\bm{e}\paral\bm{h}$ fields, it immediately follows that 
the Poynting form $\bm{\cS}$ in \fr{Poynting-2-form} vanishes, 
$\bm{\cS}=0$.    
The notation $\cdot\paral\cdot$ 
in Definition \ref{definition-Maxwell-parallel} 
is reasonable if $\bm{e}$ and $\bm{h}$  
are identified with vector fields obtained by the metric dual of $\bm{e}$
and $\bm{h}$.
\begin{Remark} 
\label{remark-Beltrami-parallel}
The set of $\bm{e}\paral\bm{h}$ fields is denoted by $\cP$, 
and the set of Beltrami-Maxwell fields is denoted by $\cB$. 
Then as shown below, it follows that $\cP\neq \cB$ and 
$\cP\cap\cB\neq\emptyset$.
\end{Remark}  
\begin{enumerate}
\item
The following is a class of $\bm{e}\paral\bm{h}$ fields 
that are also Beltrami-Maxwell fields.  
In \cite{Uehara1989}, 
a way to generate $\bm{e}\paral\bm{h}$ has been shown 
in the Gibbs' vector form. 
This is written in the form language
with Proposition\,\ref{fact-stable-Hamiltonian-structure-Maxwell-fixed-time} 
as follows. 
Consider the case that $F_{0}$ is a Beltrami-Maxwell field generated by 
a common Beltrami $1$-form $\bm{v}$ with
$$
\bm{e}
=f^{e}(t)\bm{v},\qquad
\bm{h}
=f^{h}(t)\bm{v},\qquad  
\star_{3}\bm{\dr}\bm{v}
=k\bm{v},
$$
where $k\neq 0$ is constant, $f^{e}$ and $f^{h}$ are some functions of $t$. 
In this case it follows from \fr{Maxwell-decomposed-vacuo} that 
$$
\frac{\dr f^{h}}{\dr t}
=-\frac{k}{\mu_{0}}f^{e},\qquad\text{and}\qquad
\frac{\dr f^{e}}{\dr t}
=\frac{k}{\varepsilon_{0}}f^{h}.  
$$
Solving this system, 
one has that $f^{e}$ and $f^{h}$ are trigonometric functions. 
This class of Maxwell fields is focused later. 

\item
The following is a class of $\bm{e}\paral\bm{h}$ fields that are not
Beltrami-Maxwell fields.
Put
$$
\bm{e}
=e_{0}\bm{w},\qquad
\bm{h}
=h_{0}\bm{w},\qquad
\bm{w}=c_{1}\dr x^{1}+c_{2}\dr x^{2}+c_{3}\dr x^{3},   
$$
where $e_{0}>0,h_{0}>0$ are constant, and $c_{1},c_{2},c_{3}$ are constant.
Note that $\bm{w}$ is not a rotational Beltrami $1$-form, due to
$\star_{3}\bm{\dr}\bm{w}-f\bm{w}\neq 0$. 
These fields are time-independent $\bm{e}\paral\bm{h}$ fields, 
and these are not Beltrami-Maxwell fields.
When $c_{1}=1$ and $c_{2}=c_{3}=0$, the field coincides with the one given in 
Example\,\ref{example-3-torus-SHS}.  
Meanwhile there are some time-dependent fields. 
To see these fields explicitly, put 
$$
\bm{e}=e_{0}f_{3}(x^{3})\bm{w},\qquad
\bm{h}=\frac{\varepsilon_{0}c_{0}}{k}e_{0}f_{3}^{\prime}(x^{3})\bm{w},
\qquad
\bm{w}=f_{01}(t)\dr x^{1}+f_{02}(t)\dr x^{2},
$$
where $e_{0}$ is constant, $f_{3}$ is a function of $x^{3}$,
$f_{3}^{\prime}=\dr f_{3}/\dr x^{3}$,  $f_{01}$ and $f_{02}$
are functions of $t$ such that
$$
\frac{\dr f_{01}}{\dr t}
=c_{0}kf_{02},\qquad
\frac{\dr f_{02}}{\dr t}
=-c_{0}kf_{01},\qquad
\frac{\dr^{2} f_{3}}{\dr t^{2}}
=-k^{2}f_{3},\qquad  
k\neq 0.
$$
Then straightforward calculations show that these fields are Maxwell fields. 
\item
The following is a class of Beltrami-Maxwell fields that are not 
$\bm{e}\paral\bm{h}$ fields.
Let $f$ be a function of $\xi\in\mbbR$ that is not identically zero and
satisfies   
$$
f^{\prime\prime}(\xi)
=-f(\xi),\qquad\text{where}\quad
f^{\prime}(\xi)
:=\frac{\dr f}{\dr \xi}(\xi),\quad
f^{\prime\prime}(\xi)
:=\frac{\dr^{2} f}{\dr \xi^{2}}(\xi).
$$
In addition,  put
$$
\bm{e}
=\bm{v}_{1},\quad
\bm{h}
=\frac{1}{c_{0}\mu_{0}}\bm{v}_{2},
$$
where
$$
\bm{v}_{1}
=f(x^{3}-x^{0})\dr x^{1}+f^{\prime}(x^{3}-x^{0})\dr x^{2},\quad
\bm{v}_{2}
=f^{\prime}(x^{3}-x^{0})\dr x^{1}+f(x^{3}-x^{0})\dr x^{2}.
$$
It follows that the Poynting form 
$$
\bm{e}\wedge\bm{h}
=\frac{1}{c_{0}\mu_{0}}
(f(x^{3}-x^{0})^{2}-(f^{\prime}(x^{3}-x^{0}))^{2})
\dr x^{1}\wedge\dr x^{2},
$$
does not identically vanish, since $f$ is not identically zero.  
Observe that
$\bm{v}_{1}$ and $\bm{v}_{2}$ are Beltrami $1$-forms with 
$$
\star_{3}\bm{\dr}\bm{v}_{1}
=\bm{v}_{1},\qquad\text{and}\qquad
\star_{3}\bm{\dr}\bm{v}_{2}
=-\,\bm{v}_{2}.  
$$
Then straightforward calculations show that these fields are Maxwell fields. 
\end{enumerate}

Item 1 above is focused, and  
the following is an explicit example of
time-dependent $\bm{e}\paral\bm{h}$ Beltrami-Maxwell field.
\begin{Lemma}
\label{fact-explicit-Maxwell-Beltrami-time-dependent}
Let $k\in\mbbR$ be a non-vanishing constant, and 
$\bm{v}\in\GamLamM{1}$ a non-singular $1$-form that satisfies 
$\ii_{\partial/\partial x^{0}}\bm{v}=0$. Furthermore, 
$\bm{v}$ is assumed to satisfy
$$
\star_{3}\bm{\dr}\bm{v}
=- k \bm{v},
\qquad\text{and}\qquad
\bm{\dr}\star_{3}\bm{v}
=0.
$$
Then, let 
$\bm{e},\bm{h},\bm{B},\bm{D}$ be such that  
\beqa
\bm{e}
&=&e_{0}\cos(kx^{0})\bm{v},\qquad
\bm{h}
=\frac{e_{0}}{c_{0}\mu_{0}}\sin(kx^{0})\bm{v},
\label{parallel-field-general-e-h}\\
\bm{D}
&=&\varepsilon_{0}\star_{3}\bm{e}
=\varepsilon_{0}e_{0}\cos(kx^{0})\star_{3}\bm{v},
\qquad
\bm{B}
=\mu_{0}\star_{3}\bm{h}
=\frac{e_{0}}{c_{0}}\sin(kx^{0})\star_{3}\bm{v},
\label{parallel-field-general-D-B}
\eeqa
where $e_{0}>0$ and $k$ are constant. They form a 
time-dependent $\bm{e}\paral\bm{h}$ Beltrami-Maxwell field.
\end{Lemma}
\begin{Proof}
It is shown by straightforward calculations that
these forms satisfy Maxwell's equations \fr{Maxwell-decomposed-vacuo}. 
\qed
\end{Proof}
Examples of $\bm{v}$
in Lemma\,\ref{fact-explicit-Maxwell-Beltrami-time-dependent} 
are shown in Section\,\ref{section-example-Beltrami-form}.
One choice of $\bm{v}$ is 
$$
\bm{v}
=\cos(kx^{3})\dr x^{1}+\sin(kx^{3})\dr x^{2}, 
$$
on some $\cN$.  
In this case, one has the Maxwell fields expressed as 
(4.3) and (4.4) in \cite{Uehara1989}.

As can be seen below, there are several properties of
the Maxwell fields in
Lemma\,\ref{fact-explicit-Maxwell-Beltrami-time-dependent}. 
In the following
$\bm{v}\wedge\star_{3}\bm{v}=\bm{g}^{-1}(\bm{v},\bm{v})\star_{3}1\neq 0$
is employed, which is proved 
from $\ii_{\wt{\bm{v}}^{\bm{g}}}(\bm{v}\wedge\star_{3}1)=0$ with 
$\bm{v}$ being a non-singular $1$-form. Then, 
it also follows that $\bm{v}\wedge(\star_{3}\bm{v})\wedge\dr x^{0}\neq 0$. 
\begin{enumerate}
\item    
(Non-vanishing $3$-forms).  It follows 
from $\bm{v}\wedge\star_{3}\bm{\dr}\bm{v}\neq 0$ 
that
$$
\bm{e}\wedge\bm{\dr}\bm{e}
\neq 0\qquad\text{ when}\quad \cos(kx^{0})\neq 0,
$$
and that
$$
\bm{h}\wedge\bm{\dr}\bm{h}\neq 0\qquad\text{ when}\quad
\sin(kx^{0})\neq 0. 
$$
\item
(Non-vanishing $4$-forms).  
It follows that $F_{0}$ and $F_{1}$ are non-degenerate on $\cM$. 
To verify these, $F_{0}$ and $F_{1}$ are written as   
\beqa
F_{0}&=&-c_{0}\bm{B}-\bm{e}\wedge\dr x^{0}
=-e_{0}\sin(kx^{0})\star_{3}\bm{v}-e_{0}\cos(kx^{0}) \bm{v}\wedge\dr x^{0},
\non\\
F_{1}&=&\bm{D}-c_{0}^{-1}\bm{h}\wedge\dr x^{0}
=\varepsilon_{0}e_{0}\cos(kx^{0})\star_{3}\bm{v}
-\varepsilon_{0} e_{0}\sin(kx^{0})\bm{v}\wedge\dr x^{0}
=-\varepsilon_{0}\star F_{0}.
\non
\eeqa
Then, one has on $\cM$ that
\beq
F_{0}\wedge F_{0}
=2e_{0}^{2}\sin(kx^{0})\cos(kx^{0})\bm{v}\wedge(\star_{3}\bm{v})\wedge\dr x^{0},
\quad\text{and}\quad
\bm{e}\wedge\bm{B}
=\frac{e_{0}^{2}}{c_{0}}\sin(kx^{0})\cos(kx^{0})\bm{v}\wedge\star_{3}\bm{v}.
\non
\eeq
Hence the $4$-form $F_{0}\wedge F_{0}$ and the $3$-form $\bm{e}\wedge\bm{B}$ 
do not vanish except for $\sin(kx^{0})\cos(kx^{0})=0$. Similarly
on $\cM$ it follows that 
\beq
F_{1}\wedge F_{1}
=-2e_{0}^{2}\sin(kx^{0})\cos(kx^{0})\bm{v}\wedge(\star_{3}\bm{v})\wedge\dr x^{0},
\quad\text{and}\quad
c_{0}^{-1}\bm{h}\wedge\bm{D}
=\varepsilon_{0}^{2}e_{0}^{2}\sin(kx^{0})\cos(kx^{0})\bm{v}\wedge\star_{3}\bm{v}.
\non
\eeq
Hence they do not vanish except for $\sin(kx^{0})\cos(kx^{0})=0$. 
\end{enumerate}
In addition, one has the following.
\begin{Lemma}
\label{fact-explicit-Maxwell-Beltrami-time-dependent-SHS}
Fix $x^{0}$. Let $(\cN_{x^{0}},\bm{g})$ be a $3$-dimensional 
Riemannian manifold, $\bm{B}$ and $\bm{e}$
be chosen from \fr{parallel-field-general-e-h} and
\fr{parallel-field-general-D-B}. 
Then for a fixed $x^{0}$, the pair $(\bm{B},\bm{e})$ is a stable Hamiltonian
structure for $F_{0}$ on $\cN_{x^{0}}$  
when $\sin(kx^{0})\cos(kx^{0})\neq 0$.
Its Reeb vector field is given by
$$
Y_{0}
=\frac{\wt{\bm{e}}^{\bm{g}}}{\bm{g}^{-1}(\bm{e},\bm{e})}
=\frac{1}{e_{0}\cos(kx^{0})}
\frac{\wt{\bm{v}}^{\bm{g}}}{\bm{g}^{-1}(\bm{v},\bm{v})}.
$$
Similarly,  for a fixed $x^{0}$, the pair $(\bm{D},\bm{h})$
is a stable Hamiltonian  structure for $F_{1}$ on $\cN_{x^{0}}$  
when $\sin(kx^{0})\cos(kx^{0})\neq 0$.
Its Reeb vector field is given by   
$$
Y_{1}
=\frac{\wt{\bm{h}}^{\bm{g}}}{\bm{g}^{-1}(\bm{h},\bm{h})}
=\frac{c_{0}\mu_{0}}{e_{0}\sin(kx^{0})}
\frac{\wt{\bm{v}}^{\bm{g}}}{\bm{g}^{-1}(\bm{v},\bm{v})}.  
$$
In addition, $\bm{e}$ is a contact form  
when $\cos^{2}(kx^{0})\neq 0$, and $\bm{h}$ is a contact form 
when $\sin^{2}(kx^{0})\neq 0$.
Their Reeb vector fields $Y_{0}$ and $Y_{1}$ are given above. 
\end{Lemma}
\begin{Proof}
For the proof regarding the stable Hamiltonian structures,  
applying \fr{stable-Hamiltonian-structure-Maxwell-Reeb-0-1-e-h} 
with \fr{parallel-field-general-e-h} and 
\fr{parallel-field-general-D-B}, one completes the proof. 
For the proof regarding contact forms,   
a way to prove this is analogues to the proof of 
Proposition\,\ref{fact-stable-Hamiltonian-structure-Maxwell-fixed-time}.  
\qed
\end{Proof}
In Lemma\,\ref{fact-explicit-Maxwell-Beltrami-time-dependent-SHS}, 
since $(\bm{B},\bm{e})$ is a stable Hamiltonian structure, 
$\ker\bm{B}\subset \ker(\bm{\dr}\bm{e})$ holds. It follows from this that 
there exists a function $f^{B}$ 
on $\cN$ such that $\bm{\dr}\bm{e}=f^{B}\bm{B}$. 
Similarly there exists a function $f^{D}$ such that 
$\bm{\dr}\bm{h}=f^{D}\bm{D}$.  
The functions $f^{B}$ and $f^{D}$ are found to be 
$$
f^{B}
=-\frac{c_{0}k}{\tan(kx^{0})}, \quad\text{and}\quad
f^{D}
=-c_{0}k \tan(kx^{0}),
$$
respectively.

The discussions so far are summarized as follows.
\begin{Theorem}
\label{fact-twisted-mode-geometry-summery}
The Maxwell fields 
\fr{parallel-field-general-e-h} and
\fr{parallel-field-general-D-B} 
are Beltrami-Maxwell fields that are $\bm{e}\paral\bm{h}$ fields.
They induce several manifolds and structures at a fixed $x^{0}$. 
\begin{itemize}
\item
  $(\cN_{x^{0}},\bm{e})$ is a contact manifold when $\cos(kx^{0})\neq 0$,
\item
  $(\cN_{x^{0}},\bm{h})$ is a contact manifold when $\sin(kx^{0})\neq 0$, 
\item
  $(\bm{B},\bm{e})$ is a stable Hamiltonian structure for $F_{0}$
  when $\sin(kx^{0})\cos(kx^{0})\neq0$, 
\item
  $(\bm{D},\bm{h})$ is a stable Hamiltonian structure for $F_{1}$
  when $\sin(kx^{0})\cos(kx^{0})\neq0$,  
\item
  $(\cM_{\epsilon},F_{0})$ is a symplectic manifold,
\item
  $(\cM_{\epsilon},F_{1})$ is a symplectic manifold.    
\end{itemize}
\end{Theorem}

\subsection{Topology for field lines of Maxwell fields at a fixed time}
In this subsection the main theorems in this paper are provided. 
Before stating such theorems, recall the product manifold 
$\cM_{\epsilon}=(x^{0}-\epsilon,x^{0}+\epsilon)\times\cN_{x^{0}}$ for 
a fixed $x^{0}$. Then one has the following: 
\begin{Theorem}
\label{fact-Taubes-2007-electromagnetism}  
Fix $x^{0}$, and let $(\cN_{x^{0}},\bm{g})$  
be a closed oriented connected 
$3$-dimensional Riemannian manifold.  
Let $\bm{B},\bm{D},\bm{e},\bm{h}$ be time-dependent Maxwell fields
on $\cM_{\epsilon}$  with 
$\bm{e}\wedge\bm{\dr}\bm{e}\neq 0$ and 
$\bm{h}\wedge\bm{\dr}\bm{h}\neq 0$ on $\cN_{x^{0}}$.  
Then there are closed field lines of $\bm{e}$ and $\bm{h}$ on $\cN_{x^{0}}$. 
\end{Theorem}
\begin{Proof}
Since $\bm{e}\wedge\bm{\dr}\bm{e}\neq 0$ and 
$\bm{h}\wedge\bm{\dr}\bm{h}\neq 0$, $\bm{e}$ and $\bm{h}$
are contact forms.  
In addition, 
$\lambda_{\bm{e}}:=\bm{g}^{-1}(\bm{e},\bm{e})\bm{e}$ and 
$\lambda_{\bm{h}}:=\bm{g}^{-1}(\bm{h},\bm{h})\bm{h}$ 
are also contact forms.  
Then $(\cN_{x^{0}},\lambda_{\bm{e}})$ and 
$(\cN_{x^{0}},\lambda_{\bm{h}})$   
are contact manifolds, and their Reeb vector fields are 
$$
Z_{0}
=\wt{\bm{e}}^{\bm{g}},\qquad\text{and}\qquad
Z_{1}
=\wt{\bm{h}}^{\bm{g}}.  
$$
These mean that 
the integral curves of $Z_{0}$ and $Z_{1}$ are field lines of 
$\bm{e}$ and $\bm{h}$. 
From Theorem\,\ref{fact-Taubes-2007} it follows that the integral curves of  
$Z_{0}$ and $Z_{1}$ have closed orbits. Hence the 
field lines of $\bm{e}$ and $\bm{h}$ have closed orbits.
\qed
\end{Proof}

Theorem\,\ref{fact-Taubes-2007-electromagnetism} 
is based on contact manifolds. Meanwhile 
the following is based on stable Hamiltonian structures. 
Applying Theorem\,\ref{fact-Hutchings2009}, one
immediately has the existence theorem 
for closed orbits of the electromagnetic fields $\bm{e}$ and $\bm{h}$ 
at $x^{0}=\const$. 
\begin{Theorem}
\label{fact-Hutchings2009-electromagnetism}
Fix $x^{0}$, and let $\bm{B},\bm{D},\bm{e},\bm{h}$ be time-dependent Maxwell fields on $\cM_{\epsilon}$. 
In addition, let $(\cN_{x^{0}},\bm{g})$ be a closed oriented connected 
$3$-dimensional Riemannian manifold with stable Hamiltonian
structures $(\bm{B},\bm{e}/\bm{g}^{-1}(\bm{e},\bm{e}))$ for $F_{0}$
and $(\bm{D},\bm{h}/\bm{g}^{-1}(\bm{h},\bm{h}))$ for $F_{1}$
on $\cN_{x^{0}}$. 
If $\cN_{x^{0}}$ is not a $T^{2}$-bundle over $S^{1}$, then the
field lines of $\bm{e}$ and those of $\bm{h}$ on $\cN_{x^{0}}$ 
have closed orbits. 
\end{Theorem}
\begin{Proof}
A key point in proving this statement is to identify 
the Reeb vector fields. 
The Reeb vector fields $Z_{0}$ and $Z_{1}$ are expressed as 
$$
Z_{0}
=\wt{\bm{e}}^{\bm{g}},\qquad\text{and}\qquad
Z_{1}
=\wt{\bm{h}}^{\bm{g}}.  
$$
Then applying Theorem\,\ref{fact-Hutchings2009}, one completes the proof.  
\qed 
\end{Proof}

Maxwell fields that can be applied to 
Theorems\,\ref{fact-Hutchings2009-electromagnetism}
and \ref{fact-Taubes-2007-electromagnetism} are
given in Theorem\,\ref{fact-twisted-mode-geometry-summery}. 

Physical relevance of Theorems\,\ref{fact-Taubes-2007-electromagnetism}
and \ref{fact-Hutchings2009-electromagnetism} is discussed below.
In an idealized setting for Maxwell's equations,
$\cN$ is often chosen to be $\cN=\mbbR^{3}$. 
This choice of $\cN$ does not allow us to directly apply theorems related to
the Weinstein conjecture, since the involved manifolds should be closed
in this conjecture.   
Meanwhile, the torus $T^{3}$ made from $\mbbR^{3}$ is 
often useful in numerical simulations, and 
the Maxwell systems on $T^{3}$ made from $\mbbR^{3}$ are said to have
{\it (spatially) periodic boundary conditions}. 
If a Maxwell system is defined on $\cN=T^{3}$, then
one can apply Theorems\,\ref{fact-Taubes-2007-electromagnetism}
and \ref{fact-Hutchings2009-electromagnetism}. 
This usefulness of $T^{3}$ is not limited to the study of electromagnetism,
and it is also useful in the study of fluid mechanics. 
In the study of the ideal fluid by means of contact topology, 
the following remark has been known, and this remark is also for
the study of electromagnetism.
\begin{quote}
  (Remark 4.4 of \cite{Etnyre2000}).
  \begin{it}
  The existence of closed orbits which are contractable is 
  of particular importance.
  One typically works on $T^{3}$ in order to
  model spatially periodic flows on $\mbbR^{3}$.
  The existence of a contractable closed orbit on $T^{3}$
  implies that when lifted to the universal cover $\mbbR^{3}$,
  the orbit remains closed. ... 
  \end{it}
\end{quote}
Hence the existence of contractable closed orbits in $T^{3}$
is a key ingredient for applications to Maxwell 
systems on $\mbbR^{3}$ with periodic boundary conditions. Further discussions
about this existence can be found in \cite{Etnyre2000}.  
Other physically relevant systems with closed manifolds are laser systems. 
Closed manifolds used for laser systems are called {\it cavities} or
{\it cavity resonators}.  
In cavities, the amplitudes of electromagnetic fields are amplified by 
placing mirrors. In this case one needs to verify 
that Maxwell fields satisfy appropriate boundary
conditions\,\cite{Mochizuki2022}. 

\section{Discussions and conclusions}
\label{section-dicussion-conclusions}
This paper has shown the existence of closed field lines
in closed $3$-dimensional manifolds, 
where the field lines are generated by Maxwell fields at a fixed time.
This has been summarized as main theorems, and these theorems
have been proved by applying 
the existing resolution of a basic question in dynamical systems theory,
where this question is known as the Weinstein conjecture. 
In dynamical systems theory, to determine whether or not there is 
a periodic orbit in phase space for 
a given dynamical system is one of the basic questions. 
An important class of dynamical systems is the class consisting of 
Hamiltonian systems, and a geometric description of Hamiltonian systems is 
summarized as symplectic and contact geometries.
One of the major advances in contact topology is the resolution of  
the Weinstein conjecture, where this resolution is to 
prove the existence of periodic orbits of Reeb vector fields on closed contact
manifolds. In this paper, this resolution has been applied
in electromagnetism.

Symplectic and contact geometries are active research areas and
their developments are expected to be applied to 
various branches of mathematical sciences and engineering.
On the one hand, 
contact topological methods have been applied to
the study of the ideal fluid and that of
three-body problems\,\cite{Etnyre2000,Moreno2022}.
On the other hand, other applications are less known, and  
the present study can be seen as an electromagnetic analogue of
the contact topological study of the ideal fluid.
Although the employed methods in this paper are well-known in contact topology,
the applications of such methods had never been developed in electromagnetism, 
except for a few lines of studies\,\cite{Dahl2004}. 
Hence this contribution can be seen as a step toward the
establishment of a contact topological method in 
electromagnetism. 

There remain unsolved problems that have not been addressed in this paper.
They include the following:
\begin{itemize}
\item
  generalizing the present work of Maxwell systems in vacuo to systems 
  with non-trivial media\,\cite{Dahl2013}, 

\item
  finding applications of the present study in electromagnetism,  

\item
  elucidating the role of overtwisted discs in electromagnetism,
  where overtwisted discs play important roles in contact 
  topology\,\cite{McDuff},
  and the corresponding elucidation in hydrodynamics has been
  advocated\,\cite{Etnyre2000},   

\item   
  following various works  
  on contact topological studies of the ideal fluid\,\cite{Cardona2019},
  and showing analogous theorems for electromagnetism, 

\item
  exploring the possibilities that 
  advanced methods in symplectic topology can be applied to 
  Maxwell's equations. More specifically,
  some applications of the non-squeezing theorem should be studied to reveal
  topological aspects of Maxwell's equations.
\end{itemize}

We believe that the elucidation of these remaining questions together
with this work 
will develop the geometric theories 
of electromagnetism and their applications to various related sciences.

\subsection*{Acknowledgment}
The author would like to thank Minoru Koga
for giving numerous illuminating suggestions,
fruitful discussions on this study, and 
indicating various errors on the previous version of the manuscript.  
The author also would like to thank 
Leonid Polterovich,  Yosuke Nakata, and an anonymous referee 
for  suggestions to improve the paper.  

\subsection*{Author Declarations}
\subsubsection*{Conflict of Interest}
The author has no conflicts to disclose.
\subsubsection*{Author Contributions}
Shin-itiro Goto: Conceptualization (equal);
Formal analysis (equal);
Funding acquisition (equal);
Investigation (equal);
Methodology (equal);
Project administration (equal);
Resources (equal);
Validation (equal);
Visualization (equal);
Writing – original draft (equal);
Writing – review \& editing (equal).
\subsubsection*{Data Availability}
Data sharing is not applicable to this article as no new data were created
or analyzed in this study.

\appendix 
\section{Appendix: Derivations of decomposed equations}
\label{section-appendix}
In this appendix section, details of derivations of
equations in the main text are shown. 

\subsection{Decomposed form of Maxwell's equations}
\label{section-appendix-Maxwell-decomposed}
In the following the decomposed form of Maxwell's equations
\fr{Maxwell-decomposed-vacuo} is derived. 

Substituting \fr{F0} into 
\fr{Maxwell-equations-vacuo}, one has that  
\beqa
\dr F_{0}
&=&- \dr(c_{0}\bm{B}+\bm{e}\wedge\dr x^{0})
=- c_{0}\left(\dr t\wedge\cL_{\partial/\partial t}\bm{B}
+\bm{\dr}\bm{B}\right)
-\bm{\dr}\bm{e}\wedge\dr x^{0}
\non\\
&=&-\left(\cL_{\partial/\partial t}\bm{B}+\bm{\dr}\bm{e}
\right)\wedge\dr x^{0}-c_{0}\bm{\dr}\bm{B}
=0.
\non
\eeqa
Note that $\bm{\dr}\bm{B}$ does not contain $\dr x^{0}$. 
Decomposing the above equation, $\dr F_{0}=0$,
into the term proportional to $\dr x^{0}$ and the other term, 
one obtains the first and second equations of \fr{Maxwell-decomposed-vacuo}.

Substituting \fr{F1} into 
\fr{Maxwell-equations-vacuo}, one has that
\beqa
\dr F_{1}
&=&\dr(\bm{D}-c_{0}^{-1}\bm{h}\wedge\dr x^{0})
=\dr t\wedge\cL_{\partial/\partial t}\bm{D}+\bm{\dr}\bm{D}
-c_{0}^{-1}\bm{\dr}\bm{h}\wedge\dr x^{0}
\non\\
&=&c_{0}^{-1}\left(\cL_{\partial/\partial t}\bm{D}-\bm{\dr}\bm{h}
\right)\wedge\dr x^{0}+\bm{\dr}\bm{D}
=0.
\non
\eeqa
Note that $\bm{\dr}\bm{D}$ does not contain $\dr x^{0}$. 
Decomposing 
the above equation, $\dr F_{1}=0$,
into the term propositional to $\dr x^{0}$ and the other term, 
one obtains the last two equations of \fr{Maxwell-decomposed-vacuo}.

\subsection{Decomposed form of the constitutive relations}
\label{section-appendix-constitutive-relations-decomposed}
In the following the decomposed form of the constitutive relations in vacuo
\fr{Maxwell-constitutive-relations-vacuo-decomposed} is derived.

In vacuo, the constitutive relation $F_{1}=\star\kappa F_{0}$ 
is implemented by $\kappa=-\varepsilon_{0}\Id$ with
$\Id:\GamLamM{2}\to\GamLamM{2}$ being the identity operator 
(see \fr{constitutive-relation-vacuo-0}). This reads as 
$$
F_{1}
=-\varepsilon_{0}\star F_{0}.
$$
Substituting the decomposed form of $F_{0}$ and that of $F_{1}$
into the equation above, one has the main equation: 
$$
\bm{D}-c_{0}^{-1}\bm{h}\wedge \dr x^{0}
=\varepsilon_{0}c_{0}\star\bm{B}
+\varepsilon_{0}\star \left(\bm{e}\wedge\dr x^{0}\right).
$$
To reduce the last term in the right hand side of the equation above,
one uses
$$
\star \left(\bm{e}\wedge\dr x^{0}\right)
=\ii_{\wt{\dr x^{0}}}\star \bm{e}
=\ii_{\wt{\dr x^{0}}}\ii_{\wt{\bm{e}}}\star 1
=-\ii_{\wt{\bm{e}}}\ii_{\wt{\dr x^{0}}}\star 1
=-\ii_{\wt{\bm{e}}}\ii_{\wt{\dr x^{0}}}(\dr x^{0}\wedge\star_{3}1)
=-\ii_{\wt{\bm{e}}}(-\star_{3}1)
=\star_{3}\bm{e},
$$
where
$\ii_{\wt{\dr x^{0}}}\dr x^{0}=-1$ and $\wt{\bm{e}}^{\bm{g}}=\wt{\bm{e}}$ 
have been used. 
Hence the main
equation can be written as
$$
\bm{D}-c_{0}^{-1}\bm{h}\wedge \dr x^{0}
=\varepsilon_{0}c_{0}\star\bm{B}+\varepsilon_{0}\star_{3}\bm{e}.
$$
Recall that $\bm{D}$ does not contain $\dr x^{0}$, $\ii_{\wt{\dr x^{0}}}\bm{D}=0$.
Notice then that the term $\star \bm{B}$ 
contains $\dr x^{0}$, that is, $\ii_{\wt{\dr x^{0}}}\star\bm{B}\neq 0$, and that 
the term $\star_{3}\bm{e}$ does not contain $\dr x^{0}$. 
Extracting the terms that do not contain $\dr x^{0}$ from the main equation,  
one arrives at 
$$
\bm{D}
=\varepsilon_{0}\star_{3}\bm{e}.
$$
The remaining terms in the main equation give 
the subsidiary equation: 
$$
-\bm{h}\wedge \dr x^{0}
=\varepsilon_{0}c_{0}^{2}\star\bm{B}.
$$
To reduce the left hand side of the subsidiary equation, 
applying the formula 
$$
\star\star\alpha
=(-1)^{k+1}\alpha,\qquad\forall \alpha\in\GamLamM{k},\quad
k=0,\ldots,4 
$$
and $\wt{\bm{h}}^{\bm{g}}=\wt{\bm{h}}$, one has that 
\beqa
-\bm{h}\wedge \dr x^{0}
&=&\star\star\left(\bm{h}\wedge \dr x^{0}\right)
=\star\left(\ii_{\wt{\dr x^{0}}}\star\bm{h}\right)
=\star\left(\ii_{\wt{\dr x^{0}}}\ii_{\wt{\bm{h}}}\star1\right)
=\star\left(-\,\ii_{\wt{\bm{h}}}\ii_{\wt{\dr x^{0}}}\dr x^{0}\wedge\star_{3}1\right)
\non\\
&=&
\star\left(\ii_{\wt{\bm{h}}}\star_{3}1\right)
=\star\left(\star_{3} \bm{h}\right).
\non
\eeqa
With this and $\varepsilon_{0}c_{0}^{2}=\mu_{0}^{-1}$,
the subsidiary equation is written as
$$
\star\left(\star_{3} \bm{h}\right)
=\star\left(\mu_{0}^{-1} \bm{B}\right).
$$
One then arrives at 
$$
\bm{B}=\mu_{0}\star_3\bm{h}.
$$


\end{document}